\documentclass[12pt]{article}

\usepackage{graphicx}
\usepackage{color}
\usepackage{amsmath,amssymb,mathtools,bm,braket,accents}
\usepackage{newtxtext,newtxmath}
\usepackage[top=30truemm,bottom=30truemm,left=25truemm,right=25truemm]{geometry}
\usepackage{latexsym}
\usepackage{here}
\usepackage{cite}
\usepackage{hyperref}

\newcommand{\ex}[1]{\mathrm{e}^{#1}}
\newcommand{\fr}{\frac}

\newcommand{\ca}[1]{\mathcal{#1}}
\newcommand{\bb}[1]{\mathbb{#1}}
\newcommand{\abs}[1]{\left|#1\right|}

\newcommand{\kett}[1]{ \ket{#1}\!\rangle }

\def\tr{{\mathrm{Tr}}}

  \makeatletter
    
    \@addtoreset{equation}{section}
  \makeatother

\begin{document}

\begin{titlepage}
\thispagestyle{empty}

\begin{flushright}
KYUSHU-HET-732,
\\
RIKEN-iTHEMS-Report-26,
\\

\end{flushright}

\bigskip

\begin{center}
\noindent{{\large \textbf{
The Role of Completeness in Probing Symmetry Breaking
}}}\\
\vspace{2cm}
Yuya Kusuki ${}^{1,2,3}$, Hiroyasu Tajima ${}^{4,5}$, and Shion Yamashika ${}^{6}$
\vspace{1cm}

${}^{1}${\small \sl
Institute for Advanced Study,
Kyushu University, Fukuoka 819-0395, Japan
}

${}^{2}${\small \sl
Department of Physics,
Kyushu University, Fukuoka 819-0395, Japan
}

${}^{3}${\small \sl
RIKEN Interdisciplinary Theoretical and Mathematical Sciences (iTHEMS), \\
Wako, Saitama 351-0198, Japan
}

${}^{4}${\small \sl
Department of Informatics, Faculty of Information Science and Electrical Engineering, \\
Kyushu University, Fukuoka 819-0395, Japan
}

${}^{5}${\small \sl
Japan Science and Technology Agency (JST), FOREST, \\
4-1-8 Honcho, Kawaguchi, Saitama 332-0012, Japan
}

${}^{6}${\small \sl
Department of Engineering Science, Graduate School of Informatics and Engineering, \\
The University of Electro-Communications, Chofu, Tokyo 182-8585, Japan
}

\vskip 2em
\end{center}

\begin{abstract}
\emph{Completeness} is widely recognized in quantum information as an important property of a family of monotones, because it ensures that no information relevant to state conversion is lost.
We show that completeness is also physically important in many-body systems: an incomplete measure of symmetry breaking can miss essential features of symmetry-restoration dynamics.
We study the \emph{logarithmic characteristic function} (LCF), also known as the \emph{string order parameter}, whose full family is complete for exact i.i.d.\ pure-state conversion under symmetry-preserving operations for finite groups.
We introduce a fidelity-based extension of the LCF to mixed states and identify two concrete advantages over entanglement asymmetry (EA).
First, for a finite symmetry group $G$, EA is bounded by $\log |G|$ and can approach the same saturated value for different initial states in the thermodynamic limit, making their relaxation curves indistinguishable and preventing the identification of a discrete-symmetry Mpemba effect.
By contrast, the LCF can remain extensive, with a coefficient that depends on the initial state and time, and therefore continues to distinguish their relaxation dynamics.
Second, different LCF components can exhibit distinct relaxation and crossing behavior, including cases in which EA shows no crossing.
We demonstrate these advantages in spin-chain quenches and a single-qubit system under depolarizing noise.
We also develop a replica construction for the fidelity-based LCF in quantum field theory and derive analytical results for conformal field theory.
\end{abstract}

\end{titlepage}

\restoregeometry

\tableofcontents

\section{Introduction}
\emph{Completeness} is widely recognized in quantum information as an important property of a family of monotones.
A complete family retains all the information needed to determine the relevant state conversions, whereas an incomplete measure necessarily discards some of that information~\cite{Gour2015,Sagawa2022}.
\footnote{
In general, the notion of completeness for monotones appears in two distinct settings.
In the single-shot setting, the term \emph{complete set of monotones} refers to a set of monotones that fully determines single-copy convertibility: $\rho\to\sigma$ is achievable by a free operation if and only if $M(\rho)\geq M(\sigma)$ holds for every monotone $M$ in the set~\cite{Gour2015,Sagawa2022}.
In the i.i.d.\ setting, the terms \emph{complete monotone(s)} and \emph{complete measure(s)} refer to monotones that fully determine the convertibility between many-copy states $\rho^{\otimes n}$ and $\sigma^{\otimes m}$, together with the optimal conversion rate~\cite{Sagawa2022,Shitara2023,Yamaguchi2026}.
The latter notion is well suited to capturing macroscopic properties of a system.
It is this latter sense of completeness that we adopt throughout this paper.
}
In this work, we show that this distinction also has direct physical consequences in many-body systems: information discarded by an incomplete measure can be essential for characterizing symmetry-restoration dynamics.

We investigate this issue using the \emph{logarithmic characteristic function} (LCF) in the resource theory of asymmetry (RTA) \cite{Vaccaro2008,Gour2009,Shitara2023,Yamaguchi2026,Marvian_thesis,Gour2008,Marvian2013,Marvian2014AsymmetryStates,Marvian_distillation, Marvian2022, Yamaguchi2023,Kusuki2026}.
Let $G$ be a symmetry group represented on the Hilbert space by unitary operators $U(g)$,
with $g\in G$.
For a pure state $\ket{\psi}$, the LCF is defined by
\begin{equation}
  L_\psi(g):=-\log\abs{\braket{\psi|U(g)|\psi}},
  \label{eq:Lphi}
\end{equation}
and is also known in the many-body literature as the \emph{string order parameter} \cite{Bonsignori2023}.
For a finite symmetry group, the full family $\{L_\psi(g)\}_{g\in G}$ is complete for exact i.i.d.\ pure-state conversion under symmetry-preserving operations \cite{Shitara2023}.
This operational result motivates the central question of this work: does retaining the full LCF family also provide an advantage in many-body physics?

Applications to many-body systems naturally involve mixed states, particularly reduced density matrices.
The naive extension $-\log|\tr(\rho U(g))|$ is not an appropriate measure of symmetry breaking in mixed states, since it can diverge even when $\rho$ is symmetric.
We therefore introduce the fidelity-based LCF
\begin{equation}
  \widetilde{L}_\rho(g):=-\log F\!\bigl(\rho,\mathcal{U}_g(\rho)\bigr),\qquad
  \mathcal{U}_g(\rho):=U(g)\rho U(g)^\dagger,
\end{equation}
where $F$ is the Uhlmann fidelity.
By contrast, the relative entropy of asymmetry,
known as \emph{entanglement asymmetry} (EA) when applied to reduced density matrices \cite{Ares2022},
compares the state with its group-twirled average and returns a single scalar.\footnote{
Subsequent applications of EA include spin chains and tensor-network states \cite{Capizzi2023,Lastres2024,Capizzi2023a,Russotto2024},
CFT and holography \cite{Chen2023,Fossati2024,Fossati2024a,Kusuki2024,Benini:2024xjv},
quantum-information protocols \cite{Ares2023a,Chen2024},
and gauge, higher-form, non-Abelian, and noninvertible symmetries \cite{Florio2025,Fujimura2025,Benini2025,Benini:2025hbj}.
EA has also been extensively used to study dynamical symmetry restoration and the quantum Mpemba effect \cite{Ares2023,Murciano2023,Ferro2023,Ares2024,Rylands2023,Yamashika2024,Turkeshi2024,Liu2024,Liu2024a,Chalas2024,Caceffo2024,Rylands2024,Yamashika2024a,Joshi2024,DiGiulio:2025ems,yamashika2026_long_range};
see Ref.~\cite{Ares:2025onj} for a recent review.
}
Although the completeness of the LCF family has been established only for exact i.i.d.\ conversion between pure states,
we show that retaining its individual components remains physically important for mixed-state dynamics in many-body physics.

\begin{figure}[t]
  \centering
  \includegraphics[width=0.9\linewidth]{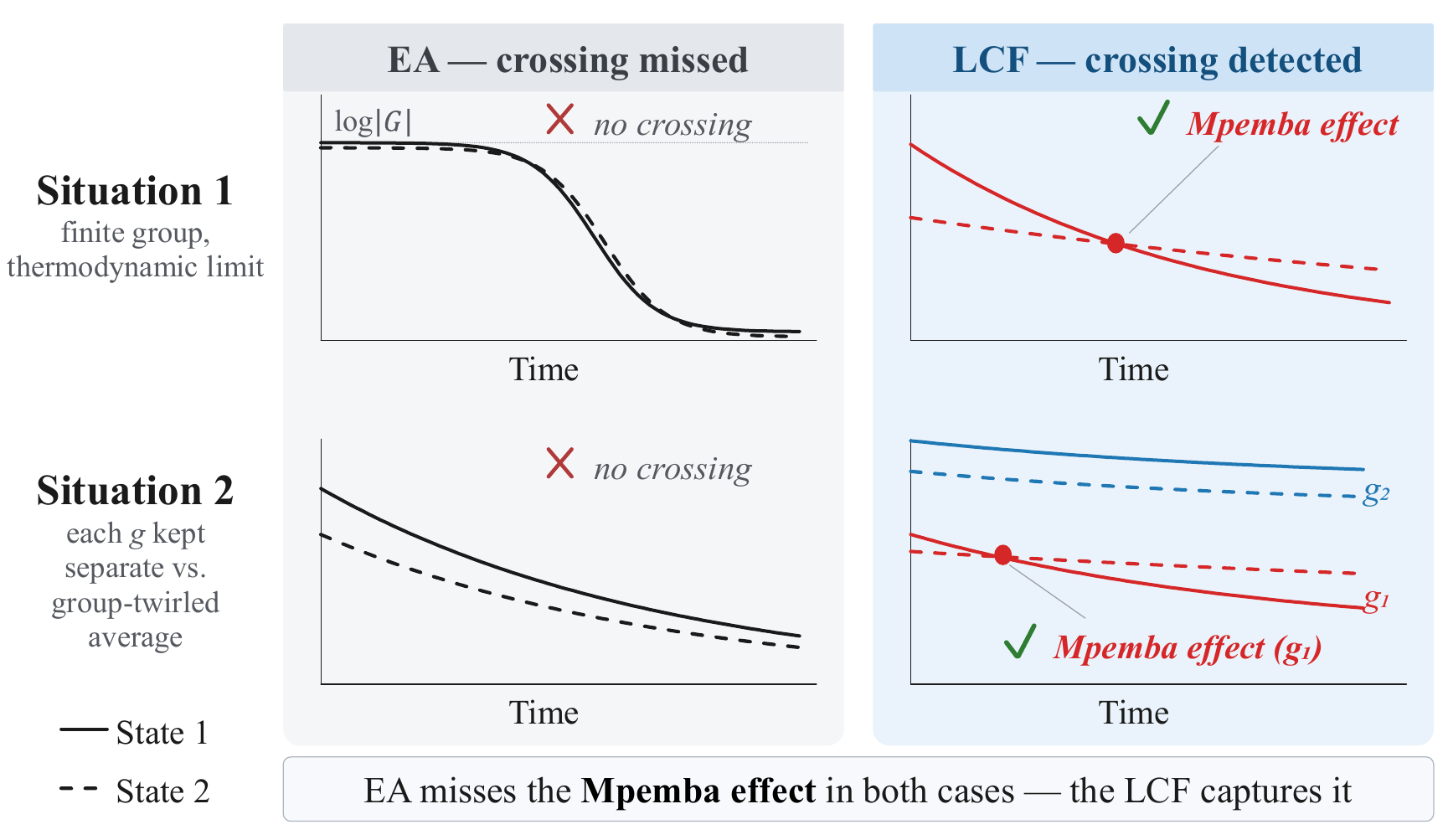}
  \caption{
    Why completeness matters for the quantum Mpemba effect.
    The complete LCF family retains the information that the scalar EA discards in two distinct ways.
    Top row, finite-group saturation in the thermodynamic limit:
    the EA curves for State~1 (solid) and State~2 (dashed) approach the same value $\log\abs{G}$ and lose the ordering needed to identify a crossing,
    whereas the extensive LCF retains the distinction between the two states and reveals a crossing at a finite time.
    Bottom row, averaging over group elements:
    EA combines the responses to different symmetry transformations into a single number and can therefore hide a crossing that occurs only for a particular group element.
    The LCF retains each component separately and detects the crossing in the $g_1$ component,
    while the $g_2$ components remain ordered.
    The family structure motivated by completeness therefore resolves both the loss of thermodynamic contrast and the loss of information under group averaging.}
  \label{fig:summary}
\end{figure}

We identify two distinct limitations of EA that can hide a crossing,
as summarized schematically in Fig.~\ref{fig:summary}.
First, for a finite symmetry group $G$, EA satisfies
\begin{equation}
  A_G(\rho)\leq \log|G|,
\end{equation}
and therefore cannot contain a contribution proportional to the subsystem size.
As pointed out for the $\bb{Z}_2$ spin-parity symmetry in the XY chain \cite{Ferro2023},
this saturation can remove the initial-state ordering needed to identify a \emph{Mpemba effect},
in which the symmetry restoration occurs more rapidly when the initial state exhibits a higher degree of symmetry breaking, in the large-subsystem limit. 
In the quench dynamics of one-dimensional cluster model studied below,
this limitation becomes decisive:
at fixed time, EA approaches the same value $\log 2$ for different initial states as the subsystem size tends to infinity,
so it cannot resolve a discrete-symmetry Mpemba effect.
The LCF, by contrast, remains extensive with a coefficient that depends on the initial state and time,
and consequently continues to distinguish the two relaxation dynamics.

Retaining the distinctions between relaxation dynamics in the large-subsystem limit is important not only for detecting the Mpemba effect but also for understanding its physical origin.
In a broad class of integrable systems, symmetry restoration can be understood through the quasiparticle picture, a semiclassical theory of quench dynamics that emerges at large subsystem sizes.
As we show in Sec.~\ref{sec:cluster_mpemba}, the fidelity-based LCF allows us to interpret the Mpemba effect for a finite-group symmetry within this picture, in parallel with its counterpart for continuous $U(1)$ symmetry~\cite{Rylands2024}.

The second limitation is independent of this finite-group saturation.
For each $g$, the corresponding LCF component compares the state with $\mathcal{U}_g(\rho)$,
whereas EA compares the state with its group-twirled average and returns a single number.
In the XXX chain, one LCF component changes ordering even though EA does not.
An analytically solvable qubit provides a stronger example,
in which an LCF component exhibits a Mpemba crossing that is absent from EA.
Thus, averaging over the group can hide differences among the individual LCF components,
including crossings that occur only in particular components.

These results make the physical significance of completeness concrete.
The motivation for complete monotones is not limited to deciding state convertibility: retaining information obscured by an incomplete scalar measure can help distinguish many-body relaxation processes and identify the microscopic origin of their differences.
Related applications of completeness in many-body physics have recently been explored in Ref.~\cite{Yamashika2025}; here we establish its role specifically in diagnosing symmetry-restoration dynamics.

We also clarify the status of the trace-based LCF, or string order parameter, in Eq.~\eqref{eq:Lphi}.
Although it is not a faithful measure of ordinary symmetry breaking for mixed states, it has an operational interpretation in a resource theory tailored to \emph{strong symmetry} \cite{Kusuki2026} and provides an upper bound on the fidelity-based LCF,
\begin{equation}
  \widetilde{L}_\rho(g)
  \leq
  L_\rho(g).
\end{equation}
This connection may provide a new operationally grounded interpretation of earlier many-body studies based on the string order parameter, including Ref.~\cite{Barad2024}: the quantity they employ can be understood as probing a strong-symmetry resource, rather than ordinary asymmetry.
At the same time, the bound explains why this quantity can remain a useful heuristic diagnostic even when it does not faithfully quantify the symmetry breaking.

Finally, we address the computability of the fidelity-based LCF in quantum field theory.
In $(1+1)$-dimensional conformal field theory, we formulate a replica construction closely related to the charged moments underlying symmetry-resolved entanglement and to replica constructions for entanglement asymmetry \cite{Kusuki2023,Kusuki2024}.
In diagonal rational conformal field theories, we show that the leading behavior of the LCF is determined by fusion coefficients and the lightest conformal weights allowed in the relevant channels.
The additional information retained by the LCF is therefore not only physically meaningful but also accessible through universal field-theory data.

The paper is organized as follows.
In Section~\ref{sec:lcf_resource_theory}, we review the RTA and the LCF for pure states.
We then introduce the fidelity-based extension to mixed states and establish its basic resource-theoretic properties.
Section~\ref{sec:qft_lcf} develops a replica construction for evaluating the fidelity-based LCF in quantum field theory.
Applying it to short-range-entangled states in two-dimensional conformal field theory (CFT),
we give an analytic example in which the LCF grows linearly with the interval length,
whereas entanglement asymmetry remains bounded for a finite symmetry group.
Section~\ref{sec:xxz_mpemba} benchmarks the fidelity-based LCF in representative XXZ quenches for which EA already exhibits a quantum Mpemba effect.
Section~\ref{sec:cluster_mpemba} studies the quench dynamics of the one-dimensional cluster model in which EA approaches the same saturated value for different initial states,
and therefore cannot resolve a discrete-symmetry Mpemba effect in the thermodynamic limit.
The LCF instead remains extensive with a coefficient that depends on the initial state and time,
allowing it to distinguish the relaxation curves and detect their crossing.
Section~\ref{sec:su2_mps} studies an $SU(2)$-symmetric XXX chain through the quaternion subgroup $Q_8\subset SU(2)$ and exhibits a crossing in one LCF component that is absent from EA.
Section~\ref{sec:depolarizing_qubit} gives an analytically solvable open-system example in which an LCF component exhibits a Mpemba crossing that EA misses.

\section{Logarithmic Characteristic Function}
\label{sec:lcf_resource_theory}

In this section, we formulate the logarithmic characteristic function (LCF) within the resource theory of asymmetry (RTA) and clarify its operational meaning as a measure of symmetry breaking.
We first review the RTA framework (see Ref.~\cite{Kusuki2026} for a review aimed at the hep-th community) and show that the quantity called the string order parameter in many-body physics is identical to the LCF for a pure state.
For pure states and finite groups, the LCF family $\{L_\psi(g)\}_{g\in G}$ is faithful and determines the exact i.i.d.\ conversion rate.
For mixed states, the direct extension $-\log|\tr(\rho U(g))|$ can be nonzero or divergent even when $\rho$ is symmetric.
We therefore construct a mixed state extension, \emph{fidelity-based LCF}, that resolves the failure of the naive trace-based definition.

\subsection{Resource theory of asymmetry}
Let $G$ be a symmetry group acting on a Hilbert space $\mathcal{H}$ through a unitary representation
$U:G\to \mathsf{U}(\mathcal{H})$.
The action on density operators $\rho \in D(\ca{H})$ is
\begin{equation}
\label{eq:group_action_rho}
  \rho \;\longmapsto\; \mathcal{U}_g(\rho) := U(g)\,\rho\,U(g)^\dagger,
  \qquad g\in G.
\end{equation}
In the resource theory of asymmetry (RTA), the set of \emph{free states} is given by the $G$-invariant states
\begin{equation}
  \label{eq:symmetric_states}
  \ca{F}
  := \bigl\{\rho \ \big|\ \mathcal{U}_g(\rho)=\rho\ \ \forall g\in G \bigr\}.
\end{equation}
The set of \emph{free operations} in the RTA consists of $G$-covariant quantum channels,
namely CPTP maps $\Lambda$ satisfying
\begin{equation}
  \label{eq:covariant_channel}
  \Lambda\circ \mathcal{U}_g \;=\; \mathcal{U}_g \circ \Lambda,
  \qquad \forall g\in G.
\end{equation}
Such operations cannot generate asymmetry from symmetric inputs; in particular,
$\rho\in \ca{F}$ implies $\Lambda(\rho)\in \ca{F}$.
We denote the set of free operations by $\ca{O}$.
This is the standard formulation of the resource theory of asymmetry \cite{Vaccaro2008,Gour2009}.

A function $M:D(\mathcal{H})\to \ca{X}$ is called an
\emph{asymmetry monotone} (or \emph{asymmetry measure}) in the RTA if it satisfies the following properties:
\begin{itemize}
  \item \textbf{Monotonicity:}
  $M$ does not increase under $G$-covariant operations,
  \begin{equation}
    M(\Lambda(\rho)) \;\le\; M(\rho),
    \qquad \forall \rho \in D(\ca{H}),\ \forall \Lambda\in\ca{O}.
    \label{eq:monotone_def}
  \end{equation}
  \item \textbf{Faithfulness:}
  $M(\rho)=0$ if and only if $\rho\in\ca{F}$.
\end{itemize}
Typically, $\ca{X}$ is taken to be $\bb{R}_{\geq 0}$.
For multi-component monotones, one may instead take $\ca{X}$ to be the set of nonnegative real-valued functions on $G$, equipped with the componentwise order.
For finite $G$, this reduces to $\bb{R}_{\geq 0}^{|G|}$, and one may equivalently represent the same order by positive semidefinite matrices.
Asymmetry monotones quantify the amount of asymmetry contained in a quantum state.
In other words, they measure the degree of symmetry breaking induced by the quantum state.

\subsection{String order parameter as multi-component resource monotone}

For a pure state $\rho = \ket{\psi}\bra{\psi}$, it is convenient to introduce the character,
\begin{equation}
  \chi_\psi(g) := \braket{\psi|U(g)|\psi}.
\end{equation}
The \emph{string order parameter} is defined as
\begin{equation}\label{eq:SOP}
  L_\psi(g) := -\log|\chi_\psi(g)|.
\end{equation}
For a fixed group element $g$, this quantity vanishes if and only if the state is invariant under that specific element,
\begin{equation}
  \label{eq:pure_lcf_elementwise}
  L_\psi(g)=0 \iff \mathcal{U}_g(\rho)=\rho .
\end{equation}
Accordingly,
\begin{equation}
  \label{eq:pure_lcf_family_faithful}
  L_\psi(g)=0\ \ \forall g\in G
  \qquad \Longleftrightarrow \qquad
  \rho\in\ca{F}.
\end{equation}
Thus the \emph{family} $\{L_\psi(g)\}_{g\in G}$ is faithful.
On this basis, \cite{Bonsignori2023} proposed the string order parameter as a suitable measure of symmetry breaking.

A key observation is that this quantity coincides exactly with the LCF \cite{Shitara2023} developed in RTA.
This identification immediately endows the string order parameter with a precise operational meaning.
For each fixed group element $g$,
$L_\psi(g)$ is non-increasing under $G$-covariant operations and detects invariance under that element.
Taken together, the collection $\{L_\psi(g)\}_{g\in G}$ therefore defines a faithful multi-component asymmetry monotone on pure states.
Moreover, for finite groups, this family has been shown to be \emph{complete} for exact i.i.d.\ conversion between pure states~\cite{Shitara2023}.
Here, exact i.i.d.\ conversion is the task of converting many copies of a state into many copies of another, with zero error, by $G$-covariant operations.
Writing $\psi^{\otimes N}\xrightarrow{\;\ca{O}\;}\phi^{\otimes M}$ when there exists $\Lambda\in\ca{O}$ such that $\Lambda(\psi^{\otimes N})=\phi^{\otimes M}$, the optimal exact conversion rate is defined as\footnote{
In the standard i.i.d.\ setting, one instead considers conversions with an error $\epsilon$ and defines the optimal rate---the approximate conversion rate---by requiring the error to vanish as $N\to\infty$.
For continuous symmetries, this standard setting indeed yields a meaningful asymptotic theory: the complete measure determining the pure-state conversion rate is the quantum Fisher information for $\mathrm{U}(1)$ symmetry~\cite{Marvian_distillation,Marvian2022,Yamaguchi2023} and the quantum geometric tensor for general compact Lie groups~\cite{Yamaguchi2026}.
For finite groups, by contrast, the approximate conversion rate either diverges or vanishes for every pair of pure states: once a vanishingly small error is allowed, asymmetry can be amplified without bound by covariant operations~\cite{Shitara2023}.
Hence the approximate setting yields no meaningful asymptotic theory for finite groups, and it is standard to consider exact i.i.d.\ conversion instead, following the seminal analysis of $\bb{Z}_2$ symmetry by Gour and Spekkens~\cite{Gour2008}.}
\begin{equation}
  \label{eq:exact_rate_def}
  R_{\mathrm{ex}}(\psi\to\phi)
  := \sup\Bigl\{\, r\ge 0 \;\Bigm|\;
  \psi^{\otimes N}\xrightarrow{\;\ca{O}\;}\phi^{\otimes \lfloor rN\rfloor}
  \ \text{for all sufficiently large}\ N \,\Bigr\}.
\end{equation}
The main result of Ref.~\cite{Shitara2023} states that this optimal rate is completely determined by the LCF family alone:
\begin{equation}
  \label{eq:exact_rate_formula}
  R_{\mathrm{ex}}(\psi\to\phi) = \min_{g\in G}\frac{L_\psi(g)}{L_\phi(g)},
\end{equation}
where the conventions $c/0:=\infty$ ($0\le c\le\infty$) and $\infty/\infty:=\infty$ are adopted in the extreme cases.
Consequently, both the convertibility and the optimal rate of exact i.i.d.\ conversion are determined solely by the LCF family, namely by the amplitudes $|\chi_\psi(g)|$ of the characteristic function.
Single-shot convertibility, by contrast, is characterized by the full characteristic function $\chi_\psi(g)$ including its phases (itself not a monotone)~\cite{Marvian2013}; this phase information becomes irrelevant in the i.i.d.\ limit.
In this sense, for finite groups the family $\{L_\psi(g)\}_{g\in G}$ constitutes a complete measure on pure states.

We emphasize that this completeness has been established only for exact i.i.d.\ conversion between pure states for finite groups, and does not imply that the LCF completely characterizes transformations between arbitrary mixed states or reduced density matrices.
Nevertheless, the pure-state result motivates us to retain one LCF component for each group element---that is, the full family---when extending the LCF to mixed states.
We then examine whether the LCF can distinguish relaxation curves that become indistinguishable in EA in the thermodynamic limit, and whether individual LCF components can detect crossings and ordering reversals that are absent from EA.

\subsection{Extension to mixed states}

The extension of the string order parameter for mixed states
was proposed in Ref.~\cite{Barad2024}.
Let us consider a spatial bipartition of the system into a region $A$ and its complement $\bar{A}$.
We assume that the corresponding Hilbert space factorizes as $\mathcal{H} = \mathcal{H}_A \otimes \mathcal{H}_{\bar{A}}$,
and that the symmetry action decomposes accordingly as $U = U_A \otimes U_{\bar{A}}$.
For a given density matrix $\rho$,
we define the reduced density matrix on $A$ by
$\rho_A := \tr_{\bar{A}} \rho$.
The naive mixed-state extension is then defined as follows.
\begin{equation}
  L_{\rho_A}(g)
  \;:=\;
  -\log \abs{\tr\ \bigl(\rho_A\,U_A(g)\bigr)}.
\end{equation}
This is precisely the ``naive'' mixed-state extension from
\eqref{eq:SOP} (with $\rho=\rho_A$ and $U(g)=U_A(g)$).
More generally, for an arbitrary mixed state $\rho$, one can define the quantity as follows.
\begin{equation}\label{eq:trace_based_lcf}
  L_{\rho}(g)
  \;:=\;
  -\log \abs{\tr\ \bigl(\rho\,U(g)\bigr)}.
\end{equation}
However, \eqref{eq:trace_based_lcf} fails as a resource-theoretic quantifier for general mixed states.
For example, consider a single qubit system, on which a $\bb{Z}_2$ symmetry acts nontrivially.
A typical $\bb{Z}_2$-symmetric mixed state is
\begin{equation}
  \rho=\frac12\Bigl(\ket{0}\!\bra{0}+\ket{1}\!\bra{1}\Bigr).
\end{equation}
For the nontrivial group element $g\in\bb{Z}_2$,
one finds
\begin{equation}
  \abs{\tr(\rho\,U(g))}=0,
\end{equation}
and hence $L_\rho(g)=+\infty$,
even though $\rho$ is fully symmetric.

To obtain a mixed-state extension compatible with RTA, we use the Uhlmann fidelity
\begin{equation}
  \label{eq:fidelity_def}
  F(\rho,\sigma)
  := \tr \sqrt{\sqrt{\rho}\,\sigma\,\sqrt{\rho}},
\end{equation}
and define the \emph{fidelity-based LCF} by
\begin{equation}
  \label{eq:mixed_lcf_def}
  \widetilde{L}_\rho(g)
  \;:=\;
  -\log F\!\bigl(\rho,\mathcal{U}_g(\rho)\bigr).
\end{equation}
This reduces to the \emph{trace-based} LCF (\ref{eq:SOP}):
if $\rho=\ket{\psi}\!\bra{\psi}$, then
$F(\rho,\mathcal{U}_g(\rho))=|\braket{\psi|U(g)|\psi}|$ and hence
$\widetilde{L}_\rho(g)=-\log|\braket{\psi|U(g)|\psi}|$.
Accordingly, \eqref{eq:mixed_lcf_def} provides the natural mixed-state extension of the string order parameter, suited to reduced density matrices.

It is worth noting that the naive extension \eqref{eq:trace_based_lcf} nevertheless admits an operational interpretation in a resource theory tailored to \emph{strong symmetry} \cite{Kusuki2026} (see Appendix \ref{app:strong}).
This viewpoint also explains why the trace-based LCF \eqref{eq:trace_based_lcf} is typically larger than the fidelity-based LCF \eqref{eq:mixed_lcf_def}.
Indeed, using
$F(\rho,\mathcal{U}_g(\rho))=\|\sqrt{\rho}\,U(g)\,\sqrt{\rho}\|_1$
and $\abs{\tr(X)}\le \|X\|_1$,
we obtain for any $\rho$ and $g$,
\begin{equation}
  \abs{\tr\ \bigl(\rho\,U(g)\bigr)}
  \;=\;
  \abs{\tr\ \bigl(\sqrt{\rho}\,U(g)\,\sqrt{\rho}\bigr)}
  \;\le\;
  \bigl\|\sqrt{\rho}\,U(g)\,\sqrt{\rho}\bigr\|_1
  \;=\;
  F\!\bigl(\rho,\mathcal{U}_g(\rho)\bigr),
\end{equation}
and therefore
\begin{equation}
-\log F\ \bigl(\rho,\mathcal{U}_g(\rho)\bigr)
  \;\le\;
  -\log \abs{\tr\ \bigl(\rho\,U(g)\bigr)},
\end{equation}
equivalently,
\begin{equation}
\widetilde{L}_\rho(g)
  \;\le\;
L_\rho(g).
\label{eq:LCFbound}
\end{equation}
Here $\|X\|_1:=\tr\sqrt{X^\dagger X}$ denotes the trace norm.
In this way, the naive extension of the string order parameter proposed in Ref.~\cite{Barad2024}
admits an operational interpretation as a measure of symmetry breaking in the strong-symmetry setting.
Moreover, even in the case of weak symmetry breaking, it provides an upper bound on the corresponding asymmetry measure.
Hence, it may constitute a useful heuristic diagnostic, though not a faithful monotone.

\subsection{Fidelity-based LCF as an asymmetry monotone}

We now verify that the fidelity-based LCF \eqref{eq:mixed_lcf_def} indeed defines a natural multi-component asymmetry monotone in the sense of the RTA \cite{Kusuki2026}.
More precisely, each fixed component $\widetilde{L}_\rho(g)$ is non-negative, monotone under free ($G$-covariant) operations, convex under mixing, and additive on product states, while the full family $\{\widetilde{L}_\rho(g)\}_{g\in G}$ is faithful.
These properties ensure that the LCF provides an operationally meaningful quantification of symmetry breaking,
placing the (generalized) string order parameter on firm resource-theoretic grounds.

\begin{description}

\item[Non-negativity]\mbox{}\\
As required for any asymmetry measure, the LCF is non-negative:
\begin{equation}
  \widetilde{L}_\rho(g) \geq 0, \qquad \forall g \in G,
\end{equation}
because $0\le F(\rho,\sigma)\le 1$.

\item[Faithfulness]\mbox{}\\
For a fixed group element $g$, the LCF vanishes if and only if the state is invariant under the action of $g$:
\begin{equation}
  \label{eq:lcf_faithful_elementwise}
  \widetilde{L}_\rho(g)=0 \iff \mathcal{U}_g(\rho)=\rho.
\end{equation}
Consequently,
\begin{equation}
  \label{eq:lcf_faithful_family}
  \widetilde{L}_\rho(g)=0\ \ \forall g\in G
  \iff
  \rho\in\ca{F}.
\end{equation}
This property ensures that the LCF family faithfully detects symmetry breaking.

\item[Monotonicity]\mbox{}\\
The requirement for a resource monotone is monotonicity under free operations.
For any $\Lambda\in\ca{O}$, the LCF satisfies
\begin{equation}
  \label{eq:lcf_monotonicity}
  \widetilde{L}_{\Lambda(\rho)}(g) \;\le\; \widetilde{L}_\rho(g),
  \qquad \forall g\in G,
\end{equation}
which follows from the monotonicity of the fidelity under CPTP maps.
Hence, asymmetry quantified by the LCF cannot be generated by symmetry-preserving processing,
which ensures that the LCF properly \emph
{measures} symmetry breaking.

\item[Convexity]\mbox{}\\
Classical mixing cannot increase asymmetry.
For an ensemble $\bar{\rho}=\sum_i p_i\rho_i$, one finds
\begin{equation}
  \label{eq:lcf_convexity}
  \widetilde{L}_{\bar{\rho}}(g)
  \;\le\;
  \sum_i p_i\, \widetilde{L}_{\rho_i}(g),
  \qquad \forall g\in G.
\end{equation}
This convexity reflects the intuition that forgetting classical information
about which state was prepared cannot enhance the asymmetry resource.

\item[Additivity]\mbox{}\\
For independent systems, asymmetry resources are additive.
Consider two systems $A$ and $B$ with representations
$U_A(g)$ and $U_B(g)$, and the product action
$U_{AB}(g)=U_A(g)\otimes U_B(g)$.
For product states $\rho_{AB}=\rho_A\otimes\rho_B$, the LCF satisfies
\begin{equation}
  \widetilde{L}_{\rho_A\otimes\rho_B}(g)
  \;=\;
  \widetilde{L}_{\rho_A}(g)+\widetilde{L}_{\rho_B}(g), \qquad \forall g\in G.
\end{equation}

\end{description}
The replacement of the trace overlap by fidelity thus changes more than the numerical behavior of the quantity.
It turns an expression that can diverge on a symmetric state into quantities that are monotone under all $G$-covariant operations and that, when considered for all $g$, detect symmetry breaking exactly.
At the same time,
exact additivity makes an extensive contribution possible rather than excluding it by construction.
They provide the resource-theoretic foundation required for using the fidelity-based LCF to compare symmetry-restoration dynamics in the remainder of this work.

\section{LCF in Quantum Field Theory}
\label{sec:qft_lcf}

To make the fidelity-based LCF accessible in quantum field theory,
we express the Uhlmann fidelity as an analytic continuation of integer replica moments that admit a Euclidean path-integral representation.
We then apply this construction to short-range entangled states in two-dimensional conformal field theory (CFT) and determine the large-interval behavior of the LCF.
We find that the LCF can grow linearly with the interval length.
For a finite symmetry group,
this contrasts with entanglement asymmetry,
whose density necessarily vanishes because it is bounded by $\log|G|$.

\subsection{Replica construction}
\label{sec:replica_lcf}

\begin{figure}[t]
  \centering
  \includegraphics[width=0.4\linewidth]{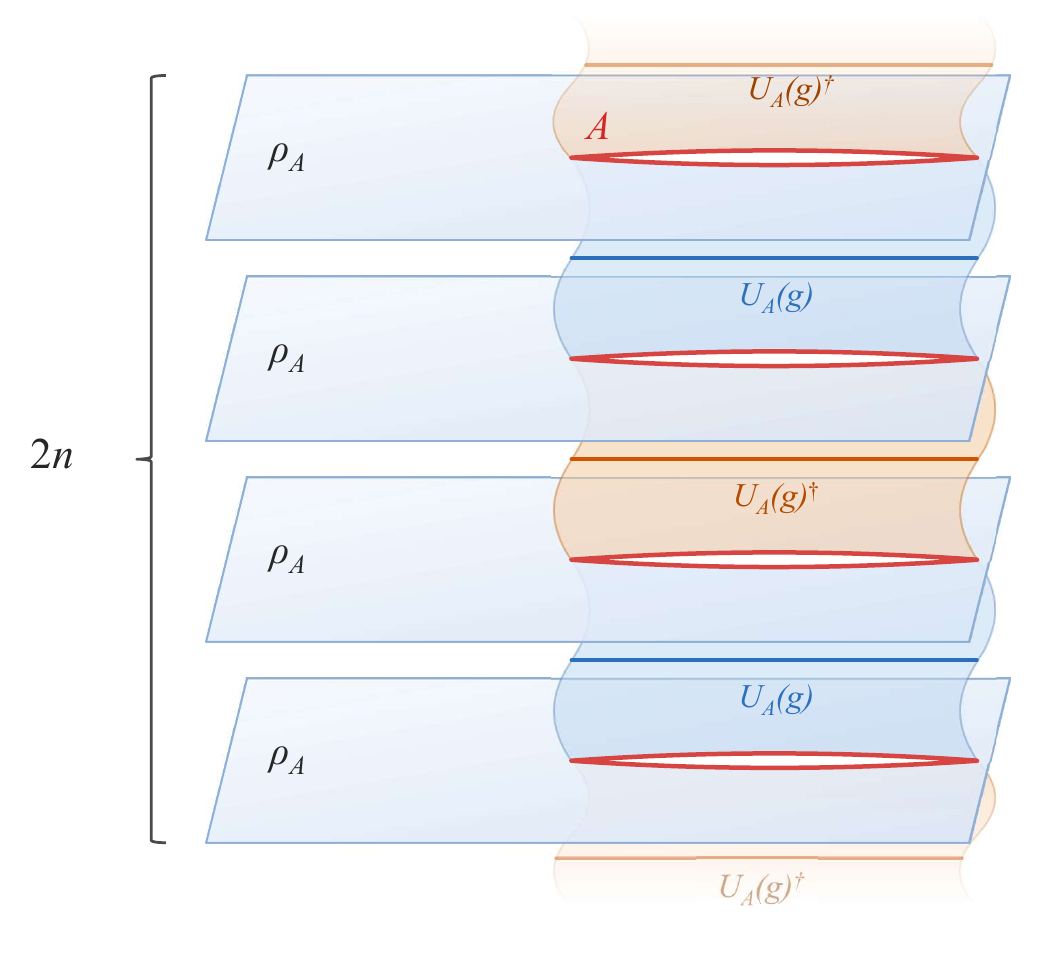}
  \caption{
    Replica manifold for $\mathcal{Z}_n(g)=\tr\!\bigl[\bigl(\rho_A U_A(g)\rho_A U_A(g)^\dagger\bigr)^n\bigr]$.
    The $2n$ copies of $\rho_A$ are glued cyclically along the cut defining $A$,
    with $U_A(g)$ and $U_A(g)^\dagger$ inserted alternately as defect lines at the gluing surfaces.
    The figure shows the case $n=2$.
  }
  \label{fig:replica}
\end{figure}

Consider a bipartition of the system into a region $A$ and its complement $\bar A$.
We assume that the Hilbert space and the symmetry action factorize as
\begin{equation}
  \mathcal{H}
  =
  \mathcal{H}_A\otimes\mathcal{H}_{\bar A},
  \qquad
  U(g)
  =
  U_A(g)\otimes U_{\bar A}(g),
  \qquad
  g\in G.
\end{equation}
For a density matrix $\rho$ on $\mathcal{H}$,
let $\rho_A:=\tr_{\bar A}\rho$.
We then define
\begin{equation}
  \label{eq:lcf_def_again}
  \sigma_A(g)
  :=
  U_A(g)\rho_AU_A(g)^\dagger,
  \qquad
  \widetilde{L}_A(g)
  :=
  -\log F\!\bigl(\rho_A,\sigma_A(g)\bigr).
\end{equation}
For a positive integer $n$, introduce the replica moment
\begin{equation}
  \mathcal{Z}_n(g)
  :=
  \tr\!\left[
  \bigl(\sqrt{\rho_A}\,\sigma_A(g)\,\sqrt{\rho_A}\bigr)^n
  \right],
  \qquad
  n\in\mathbb{N}.
\end{equation}
Cyclicity of the trace gives
\begin{align}
  \mathcal{Z}_n(g)
  &=
  \tr\!\left[
  \bigl(\rho_A\sigma_A(g)\bigr)^n
  \right]
  \nonumber\\
  &=
  \tr\!\left[
  \bigl(\rho_AU_A(g)\rho_AU_A(g)^\dagger\bigr)^n
  \right].
  \label{eq:replica_moment_product}
\end{align}
The final expression contains $2n$ copies of the reduced density matrix,
with $U_A(g)$ and $U_A(g)^\dagger$ inserted alternately between neighboring copies.
In a Euclidean path integral, these symmetry operators become defect lines along the cut defining $A$,
so that $\mathcal{Z}_n(g)$ is represented by a $2n$-sheeted replica geometry.
The gluing pattern is illustrated for $n=2$ in Fig.~\ref{fig:replica}.
This construction is closely related to the charged moments underlying symmetry-resolved entanglement and to replica constructions for entanglement asymmetry \cite{Kusuki2023,Kusuki2024},
but the fidelity is selected by the continuation $n\to\frac12$ rather than the von Neumann limit.
The Uhlmann fidelity is then recovered as
\begin{equation}
  F\!\bigl(\rho_A,\sigma_A(g)\bigr)
  =
  \lim_{n\to\frac12}\mathcal{Z}_n(g),
\end{equation}
and hence
\begin{equation}
  \label{eq:L_from_Z}
  \widetilde{L}_A(g)
  =
  -\lim_{n\to\frac12}\log\mathcal{Z}_n(g).
\end{equation}
For the identity element,
$\mathcal{Z}_n(e)=\tr(\rho_A^{2n})$,
so that $\lim_{n\to\frac12}\mathcal{Z}_n(e)=\tr\rho_A=1$ and $\widetilde{L}_A(e)=0$.
Equation~\eqref{eq:L_from_Z} is the starting point for the CFT calculation below.

\subsection{Extensive scaling in rational conformal field theory}
\label{sec:examples}

We now apply this construction to a standard class of short-range entangled states in two-dimensional CFT,
obtained by imaginary-time regularization of conformal boundary states.
We show that the LCF can grow linearly with the interval length.

We work in a unitary diagonal rational conformal field theory (RCFT).
An elementary conformal boundary condition labeled by $b$ is represented by the Cardy state \cite{Cardy2004}
\begin{equation}
  \label{eq:Cardy_state_def}
  \ket{B_b}
  :=
  \sum_{i\in\mathcal{I}}
  \frac{S_{bi}}{\sqrt{S_{0i}}}\,\kett{i}.
\end{equation}
Here $\mathcal{I}$ labels the irreducible representations of the chiral algebra,
$S_{ij}$ is the modular $S$ matrix,
$0$ denotes the vacuum representation,
and $\kett{i}$ is the Ishibashi state associated with representation $i$,
\begin{equation}\label{eq:Ishibashi}
\kett{i} := \sum_{N} \ket{i;N} \otimes   U \overline{\ket{i;N}}\,.
\end{equation}
Here, the state $\ket{i;N}$ resides in the Verma module of the irreducible representation $i$,
with $N$ indexing individual states within this module.
The operator $U$, which acts on the anti-holomorphic sector,
is an anti-unitary operator.
For example,
in the Ising CFT the label $b$ distinguishes fixed-up, fixed-down, and free boundary conditions.

The Cardy state $\ket{B_b}$ itself is not normalizable and is not the physical short-range-entangled state considered below.
We instead use the regularized state
\begin{equation}
  \label{eq:phib}
  \ket{\phi_b}
  :=
  \frac{\ex{-\frac{\beta}{4}H}}{\sqrt{Z_b(\beta)}}\ket{B_b},
  \qquad
  Z_b(\beta)
  =
  \braket{B_b|\ex{-\frac{\beta}{2}H}|B_b},
\end{equation}
which is commonly used as a short-range entangled initial state in CFT descriptions of quantum quenches \cite{Calabrese2016}.
Here, $H$ denotes the CFT Hamiltonian, which is invariant under a symmetry group $G$.
The parameter $\beta$ sets its short-distance and correlation-length scale.

We consider a finite internal symmetry,
with $g\in G$ represented by an invertible topological line $\mathcal{L}_g$.
Its action maps the boundary condition $b$ to another elementary boundary condition,
which we denote by $g\cdot b$,
\begin{equation}
  \mathcal{L}_g\ket{B_b}=\ket{B_{g\cdot b}}.
\end{equation}
Since the symmetry commutes with the Hamiltonian,
it also maps $\ket{\phi_b}$ to $\ket{\phi_{g\cdot b}}$.
For an interval $A$ of length $\ell$, define
\begin{equation}
  \rho_A^{(b)}
  :=
  \tr_{\bar A}\ket{\phi_b}\bra{\phi_b}.
\end{equation}
Its symmetry transform is
\begin{equation}
  U_A(g)\rho_A^{(b)}U_A(g)^\dagger
  =
  \rho_A^{(g\cdot b)}.
\end{equation}
If $g\cdot b=b$, the two reduced density matrices coincide and the fidelity-based LCF vanishes exactly.
Suppose instead that $g\cdot b\neq b$.
The leading decay is determined by the lightest boundary operator that connects the two boundary conditions.
For Cardy boundaries, its conformal weight is
\begin{equation}
  \label{eq:hmin_CFT}
  h_{\mathrm{min}}
  :=
  \min\Bigl\{
  h_i
  \ \Big|\ 
  N_{i b}^{\,g\cdot b}\neq0
  \Bigr\}.
\end{equation}
Here $N_{i b}^{\,g\cdot b}$ is a fusion coefficient,
or equivalently the multiplicity of sector $i$ in the spectrum of boundary operators connecting $b$ and $g\cdot b$.
The identity operator is present when $g\cdot b=b$,
whereas $h_{\mathrm{min}}>0$ when $g$ maps $b$ to a distinct elementary boundary condition.

For $\ell\gg\beta$, the replica path integral is dominated by the lightest state that can propagate along the interval.
Using the replica construction above and the analytic continuation to $n=\frac12$ gives
\begin{equation}
  \label{eq:CFT_LCF_scaling}
  \widetilde{L}_A^{(b)}(g)
  =
  \begin{cases}
    0,
    & g\cdot b=b,\\[6pt]
    \displaystyle
    \frac{2\pi\ell}{\beta}h_{\mathrm{min}}+O(1),
    & g\cdot b\neq b,
  \end{cases}
  \qquad
  \ell\gg\beta.
\end{equation}
The derivation is given in Appendix~\ref{app:rcft_lcf}.

For a finite group, entanglement asymmetry satisfies $A_G(\rho_A)\leq\log|G|$~\cite{Ferro2023}.
Consequently, when $g\cdot b\neq b$,
\begin{equation}
  \lim_{\ell\to\infty}
  \frac{\widetilde{L}_A^{(b)}(g)}{\ell}
  =
  \frac{2\pi}{\beta}h_{\mathrm{min}}>0,
  \qquad
  \lim_{\ell\to\infty}
  \frac{A_G(\rho_A^{(b)})}{\ell}
  =
  0.
\end{equation}
Thus, the CFT calculation establishes analytically the first advantage of the LCF identified in the Introduction.
For a finite symmetry group, EA is bounded by $\log|G|$ and therefore has vanishing density in the large-interval limit,
whereas the fidelity-based LCF can remain extensive with a nonzero coefficient fixed here by the lightest allowed boundary operator.
Section~\ref{sec:cluster_mpemba} demonstrates the dynamical consequence of this difference.
In the cluster-model quench,
the extensive coefficient depends on the initial state and time,
so the LCF continues to distinguish the relaxation curves even though EA approaches the same saturated value in the thermodynamic limit.

The Ising CFT gives the simplest example.
The $\bb{Z}_2$ spin flip exchanges the fixed-up and fixed-down boundary conditions,
while leaving the free boundary condition invariant.
The lightest boundary operator connecting the two fixed boundary conditions has $h_{\mathrm{min}}=\frac12$.
Equation~\eqref{eq:CFT_LCF_scaling} therefore gives
\begin{equation}
  \widetilde{L}_A^{(+)}(g)
  =
  \widetilde{L}_A^{(-)}(g)
  =
  \frac{\pi\ell}{\beta}+O(1),
  \qquad
  \widetilde{L}_A^{(\mathrm{free})}(g)=0.
\end{equation}
The reduced density matrices prepared from the two fixed boundary conditions become exponentially distinguishable from their spin-flipped images as the interval grows,
whereas the free-boundary state is unchanged by the symmetry.

\section{Benchmark in the XXZ Spin Chain}
\label{sec:xxz_mpemba}

Before turning to cases in which the fidelity-based LCF reveals behavior absent from EA,
we test it in a standard many-body setting where the quantum Mpemba effect is already established.
We study representative XXZ quenches considered in Ref.~\cite{Ares2022},
where the effect was diagnosed using the $U(1)$ entanglement asymmetry.
Because the comparison below is formulated for the finite subgroup $\bb{Z}_4\subset U(1)$,
we first verify that the entanglement asymmetry for the $\mathbb{Z}_4$ symmetry exhibits the same Mpemba effect.
We then ask whether the two inequivalent nontrivial LCF components reproduce this crossing,
and whether the trace-based LCF proposed in Ref.~\cite{Barad2024} can serve as a practical proxy for the fidelity-based LCF.

\subsection{Benchmark of the fidelity-based LCF}
\label{subsec:xxz_mpemba}

We consider an open spin chain prepared in the tilted ferromagnetic product state
\begin{equation}
  |\Psi_{\theta}(0)\rangle
  =
  \exp\Bigl(-i\frac{\theta}{2}\sum_{j=1}^{N}\sigma_j^y\Bigr)
  |\uparrow\uparrow\cdots\uparrow\rangle,
\end{equation}
and evolved with the XXZ-type Hamiltonian
\begin{equation}
  H
  =
  -\frac14\sum_{j=1}^{N-1}
  \Bigl(
  \sigma_j^x\sigma_{j+1}^x
  +
  \sigma_j^y\sigma_{j+1}^y
  +
  \Delta\,\sigma_j^z\sigma_{j+1}^z
  \Bigr)
  -\frac{J_2}{4}\sum_{j=1}^{N-2}
  \Bigl(
  \sigma_j^x\sigma_{j+2}^x
  +
  \sigma_j^y\sigma_{j+2}^y
  +
  \Delta_2\,\sigma_j^z\sigma_{j+2}^z
  \Bigr),
\end{equation}
with open boundary conditions.
The Hamiltonian commutes with the total magnetization
\begin{equation}
  Q
  :=
  \frac12\sum_{j=1}^{N}\sigma_j^z.
\end{equation}
For $\theta\neq0,\pi$,
the initial state breaks the corresponding $U(1)$ symmetry.
The quench therefore provides a standard setting for studying local symmetry restoration.

For a subsystem $A$ of length $\ell$, define
\begin{equation}
  \rho_A(t)
  :=
  \tr_{\bar A}\Bigl(
  e^{-itH}|\Psi_{\theta}(0)\rangle\!\langle\Psi_{\theta}(0)|e^{itH}
  \Bigr),
\end{equation}
and the subsystem charge
\begin{equation}
  Q_A
  :=
  \frac12\sum_{j\in A}\sigma_j^z.
\end{equation}
As the incomplete benchmark,
we use the entanglement asymmetry for the finite subgroup $\bb{Z}_4\subset U(1)$,
\begin{equation}
  A_{\bb{Z}_4}(\rho_A(t))
  :=
  S\!\bigl(\mathcal{G}_A^{\bb{Z}_4}[\rho_A(t)]\bigr)-S(\rho_A(t)),
\end{equation}
where $S(\rho):=-\tr(\rho\log\rho)$ is the von Neumann entropy, and
\begin{equation}
  \mathcal{G}_A^{\bb{Z}_4}[\rho]
  :=
  \frac14\sum_{m=0}^{3}
  e^{i\frac{\pi m}{2}Q_A}\rho e^{-i\frac{\pi m}{2}Q_A}
\end{equation}
is the twirling map.

Following the terminology in Ref.~\cite{Fraenkel2020},
we call the $\alpha$-dependent description ``flux-resolved,''
in contrast to a projection onto fixed-charge sectors.
We define
\begin{equation}
  \widetilde{L}_A(\alpha,t)
  :=
  -\log F\!\Bigl(
  \rho_A(t),
  e^{i\alpha Q_A}\rho_A(t)e^{-i\alpha Q_A}
  \Bigr).
\end{equation}
For $\bb{Z}_4$,
the parameter $\alpha$ takes the values $0,\pi/2,\pi,3\pi/2$.
The LCF component at $\alpha=0$ vanishes,
and $\widetilde{L}_A(-\alpha,t)=\widetilde{L}_A(\alpha,t)$.
The two inequivalent nontrivial LCF components therefore correspond to $\alpha=\pi/2$ and $\alpha=\pi$.

\begin{figure}[t]
  \centering
  \includegraphics[width=\textwidth]{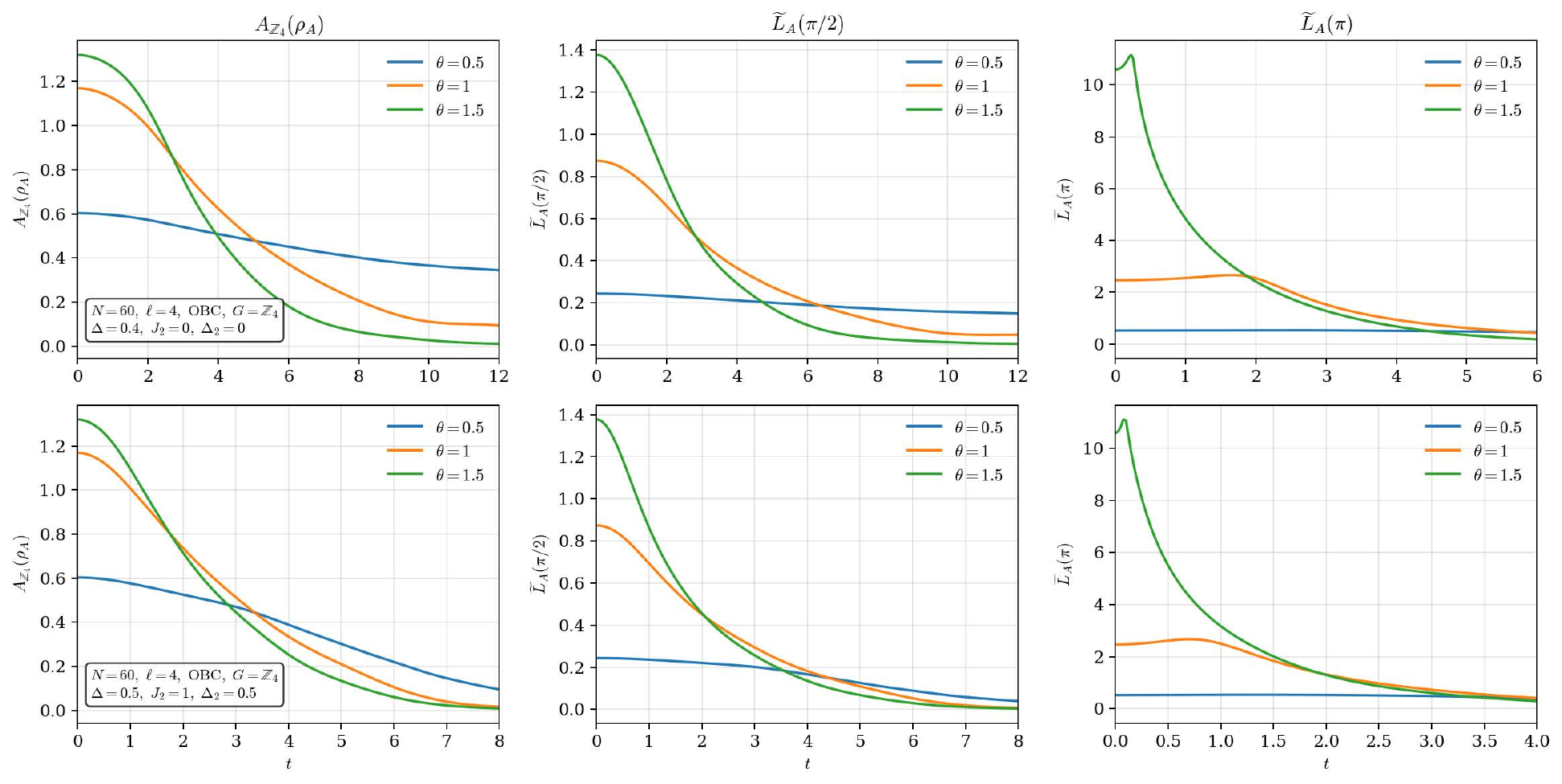}
\caption{
Entanglement asymmetry and fidelity-based LCF in two representative XXZ quenches.
All panels show numerical data for an open chain with $N=60$ and $\ell=4$, obtained using the time-evolving block decimation (TEBD) algorithm~\cite{vidal2003}. 
The TEBD algorithm is implemented by using the second-order Suzuki-Trotter decomposition with time step $\delta t=0.025$ and setting the bond dimension $\chi=128$. 
The first column shows EA,
which provides the established benchmark crossing.
The second and third columns show the two inequivalent nontrivial $\mathbb{Z}_4$ LCF components,
corresponding to $\alpha=\pi/2$ and $\alpha=\pi$.
In both examples, the LCF components also exhibit crossings,
confirming that the fidelity-based LCF captures the known Mpemba behavior.
Their crossing times and detailed time dependence need not coincide with those of EA or with each other.
}
  \label{fig:xxz_mpemba_lcf}
\end{figure}

Fig.~\ref{fig:xxz_mpemba_lcf} shows these quantities in the two representative setups for the quantum Mpemba effect. 
The main message is that the fidelity-based LCF does detect the Mpemba effect, but not in a universal or one-dimensional way.
Both $\widetilde{L}_A(\pi/2,t)$ and $\widetilde{L}_A(\pi,t)$ can show crossings, yet their crossing times need not coincide with each other or with that of $  A_{\bb{Z}_4}(\rho_A(t))$.
In particular, one LCF component can cross before $ A_{\bb{Z}_4}(\rho_A(t))$ in one setup, while in another setup $  A_{\bb{Z}_4}(\rho_A(t))$ can cross first.
The conclusion is therefore not that the LCF is simply a uniformly more sensitive scalar probe.
Rather, the LCF components at $\alpha=\pi/2$ and $\alpha=\pi$ can follow distinct relaxation curves.

This difference is naturally understood from the charge-sector decomposition.
Let $\Pi_q$ denote the projector onto the eigenspace of $Q_A$ with eigenvalue $q$.
Then
\begin{equation}
  \rho_A(t)
  =
  \sum_{q,q'}\Pi_q\rho_A(t)\Pi_{q'},
\end{equation}
and
\begin{equation}
  e^{i\alpha Q_A}\Pi_q\rho_A(t)\Pi_{q'}e^{-i\alpha Q_A}
  =
  e^{i\alpha(q-q')}\Pi_q\rho_A(t)\Pi_{q'}.
\end{equation}
Changing $\alpha$ changes the phases attached to the off-diagonal charge blocks before the fidelity is evaluated.
The resulting symmetry-transformed state therefore depends on $\alpha$,
so the LCF components at $\alpha=\pi/2$ and $\alpha=\pi$ need not cross at the same time as each other or as $A_{\bb{Z}_4}(\rho_A(t))$.
This is the additional information retained by keeping the dependence on $\alpha$ explicit.

\subsection{Comparison with trace-based LCF}
\label{subsec:xxz_trace_compare}

\begin{figure}
  \centering
  \includegraphics[width=\textwidth]{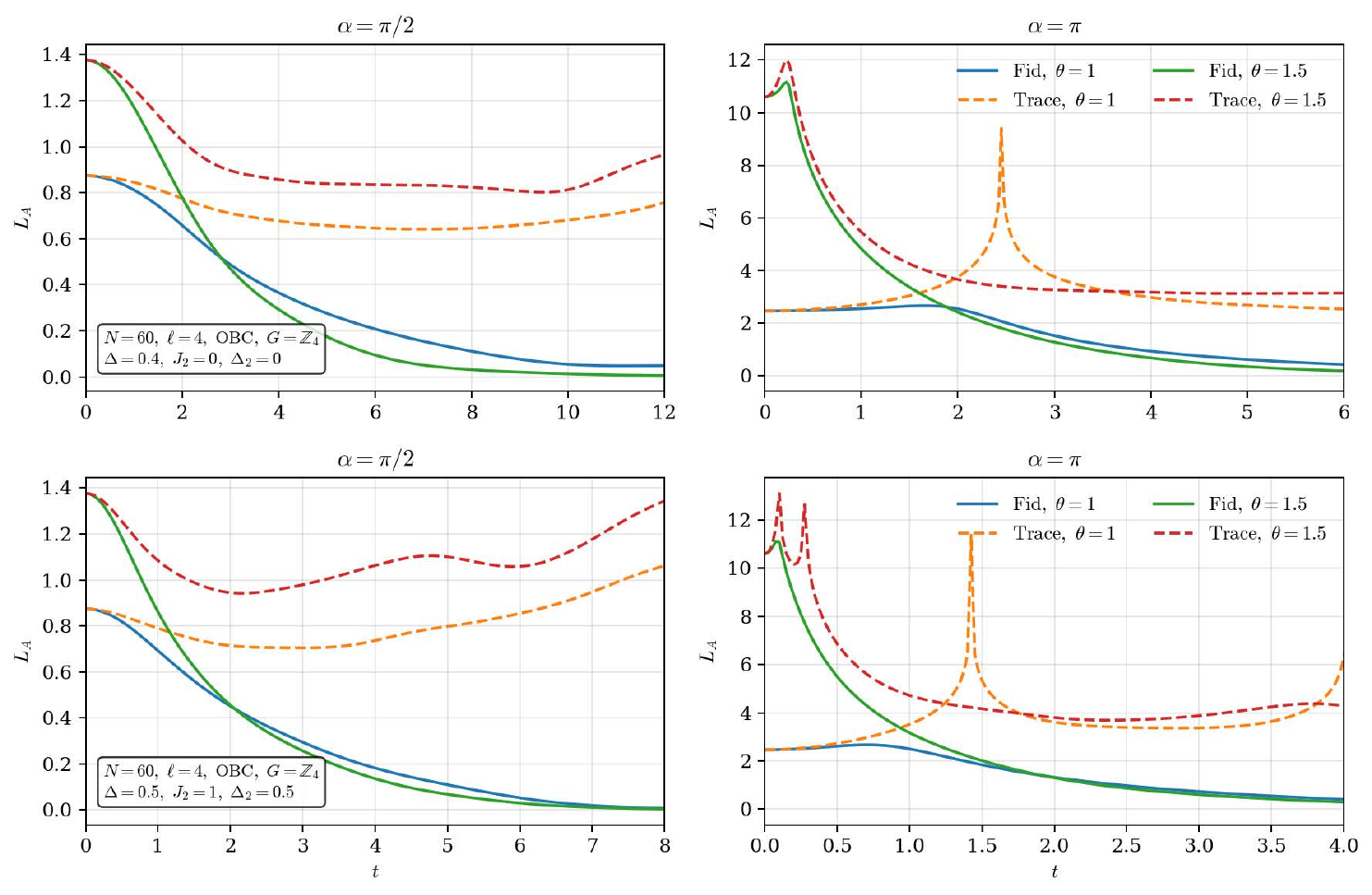}
\caption{
Fidelity-based and trace-based LCFs in the same many-body quenches.
All panels show numerical data for an open chain with $N=60$ and $\ell=4$, obtained using the TEBD algorithm.
The TEBD algorithm is implemented by using the second-order Suzuki-Trotter decomposition with time step $\delta t=0.025$ and setting the bond dimension $\chi=128$. 
Compared with the fidelity-based result, the trace-based LCF shows a weaker and more fragile Mpemba signature:
no crossing is visible at $\alpha=\pi/2$ in the plotted window, and the crossing at $\alpha=\pi$ appears only in a limited case and is accompanied by spike-like features.
These spike-like features reflect the enhanced sensitivity of the trace-based quantity to cancellations among charge sectors.
}
  \label{fig:xxz_trace_lcf}
\end{figure}

We now ask whether the same Mpemba-like behavior is also visible in the trace-based LCF,
\begin{equation}
  L_A(\alpha,t)
  :=
  -\log\left|
  \tr\!\bigl(\rho_A(t)e^{i\alpha Q_A}\bigr)
  \right|.
\end{equation}
Fig.~\ref{fig:xxz_trace_lcf} shows the corresponding time evolution.
Compared with the fidelity-based result, the trace-based LCF exhibits a much weaker and less robust Mpemba signature.
In the examples shown here,
no crossing appears at $\alpha=\pi/2$ within the plotted time windows.
At $\alpha=\pi$,
a clear crossing is visible in only one of the two quenches,
and it is accompanied by sharp spike-like features.
Thus the trace-based LCF can display a Mpemba-like crossing, but only in a restricted and rather fragile manner.

The reason is structural:
\begin{equation}
  \tr\!\bigl(\rho_A(t)e^{i\alpha Q_A}\bigr)
  =
  \sum_{q} p_{q}(t)e^{i\alpha q},
  \qquad
  p_{q}(t):=\tr\!\bigl(\Pi_{q}\rho_A(t)\bigr).
\end{equation}
The trace-based LCF depends only on the subsystem charge distribution $p_{q}(t)$ and not on the full inter-sector coherence structure of $\rho_A(t)$.
As a result, it is especially sensitive to cancellations among charge sectors, and these cancellations are what generate the spike-like behavior seen in the numerics.
By contrast, the fidelity-based LCF probes the full reduced density matrix and therefore gives a cleaner and more informative flux-resolved picture of the relaxation.

This point is important when interpreting earlier dynamical studies based on the trace-based LCF.
For example, Ref.~\cite{Barad2024} investigated symmetry breaking in a dynamical setup precisely through this quantity.
Our discussion does not invalidate such results as probes of nontrivial dynamics.
What it shows instead is that some care is needed if one wishes to interpret the trace-based LCF as a quantitative measure of \emph{how much} the symmetry is broken.
For reduced density matrices, cancellations in $\sum_q p_q(t)e^{i\alpha q}$ can make $L_A(\alpha,t)$ large or even divergent, even when $\rho_A(t)$ is symmetric.
A large trace-based LCF therefore does not necessarily imply a large fidelity-based LCF.

The numerical results in Fig.~\ref{fig:xxz_trace_lcf} are also consistent with the general bound \eqref{eq:LCFbound}, namely
\begin{equation}
  \widetilde{L}_A(\alpha,t)\le L_A(\alpha,t),
\end{equation}
which shows that the trace-based LCF provides an upper bound on the fidelity-based LCF.
From this viewpoint, the trace-based LCF is best regarded as a coarser upper-bound diagnostic, controlled mainly by charge-population imbalances and cancellations.

\section{Mpemba Effect for Finite Group Symmetry}
\label{sec:cluster_mpemba}

We now demonstrate that the extensivity of the fidelity-based LCF enables the analysis of quantum Mpemba effects associated with finite-group symmetries. We study the quench dynamics of the one-dimensional cluster model and compare the EA and the LCF for reduced density matrices of a local subsystem. In the large-subsystem limit, the EA saturates at the same value for different initial states, whereas the LCF remains extensive and distinguishes their degrees of asymmetry. We use this distinction to identify the quantum Mpemba effect and provide its microscopic interpretation based on the quasiparticle picture.

\subsection{Cluster model}
\label{subsec:cluster_setup}

We study a spin-$\frac12$ chain described by the generalized  cluster Hamiltonian:
\begin{align}
  H(J_1,J_2,\delta)
  ={}&-\frac{J_1}{4}\sum_j
  \left(
    \sigma_j^x\sigma_{j+1}^x
    +\sigma_j^y\sigma_{j+1}^y
  \right)
  \nonumber\\
  &-\frac{J_2}{4}\sum_j(-1)^j
  \left(
    \sigma_j^x\sigma_{j+1}^z\sigma_{j+2}^x
    +\sigma_j^y\sigma_{j+1}^z\sigma_{j+2}^y
  \right)
  -\frac{\delta}{2}\sum_j(-1)^j\sigma_j^z,
  \label{eq:cluster_hamiltonian}
\end{align}
where $J_1$ and $J_2$ are the strengths of the nearest-neighbor XY interaction and the staggered three-spin cluster interaction, respectively, and $\delta$ is the staggered magnetic field.
We denote the system length by $2L$ and impose periodic boundary conditions.

Under the Jordan-Wigner transformation
\begin{align}
  a_j = \frac{\sigma_{2j-1}^x-i \sigma_{2j-1}^y}{2}\prod_{l<2j-1}(-\sigma_l^z),\quad 
  b_j = \frac{\sigma_{2j}^x-i \sigma_{2j}^y}{2}\prod_{l<2j}(-\sigma_l^z),\quad (j=1,2,...,L), 
\end{align}
Eq.~\eqref{eq:cluster_hamiltonian} maps to the free-fermion Hamiltonian
\begin{equation}
  H
  =
  -\frac12\sum_i
  \left[
    J_1(a_i^\dag b_i+b_i^\dag a_{i+1})
    +J_2(a_i^\dag a_{i+1}-b_i^\dag b_{i+1})
    -\delta (a_i^\dag a_i-b_i^\dag b_i)
    +\mathrm{H.c.}
  \right].
  \label{eq:cluster_fermion_hamiltonian}
\end{equation}
After the Fourier transform $(a_k,b_k)=L^{-1/2}\sum_j e^{-ikj}(a_j,b_j)$ and the Bogoliubov transformation
\begin{align}
  \left(\begin{array}{c}
    \alpha_k \\ \beta_k 
  \end{array}\right)
  = 
  \left(
    \begin{array}{cc}
      e^{i\frac{\phi_k}{2}}\cos\frac{\theta_k}{2}
      & 
      e^{-i\frac{\phi_k}{2}}\sin\frac{\theta_k}{2}
      \\
      -e^{i\frac{\phi_k}{2}}\sin\frac{\theta_k}{2}
      &
      e^{-i\frac{\phi_k}{2}}\cos\frac{\theta_k}{2}
    \end{array}
  \right)
  \left(\begin{array}{c}
    a_k \\ b_k 
  \end{array}\right), 
\end{align}
where
\begin{align}
  \sin\theta_k\cos\phi_k=-\frac{J_1}{\omega_k}\cos^2\frac{k}{2},\quad 
  \sin\theta_k \sin\phi_k = -\frac{J_1}{2\omega_k}\sin k,\quad
  \cos \theta_k = \frac{\delta-J_2\cos k}{\omega_k},
\end{align}
the Hamiltonian in Eq.~\eqref{eq:cluster_fermion_hamiltonian} takes the diagonal form
\begin{align}\label{eq:H_cluster_diagonal}
  H = \sum_k \omega_k (\alpha_k^\dag \alpha_k-\beta_k^\dag \beta_k). 
\end{align}
Here $\omega_k=[J_1^2\cos^2\frac{k}{2}+(\delta-J_2\cos k)^2]^{1/2}$ is the dispersion relation of the Bogoliubov quasiparticles.

Equation~\eqref{eq:H_cluster_diagonal} shows that the energy spectrum consists of upper and lower bands with dispersions $\pm \omega_k$, and that $\alpha_k$ and $\beta_k$ are the quasiparticle operators associated with the upper and lower bands, respectively.
The initial ground state $\ket{\psi_0}$ has a fully occupied lower band and an empty upper band as
\begin{align}
  \ket{\psi_0}=\prod_{\forall k} \beta_k^\dag \ket{0}, 
\end{align}
where $\ket{0}$ is the fermionic vacuum state annihilated by $a_i$ and $b_i$, i.e., $a_i\ket{0}=b_i\ket{0}=0$ for all $i$. 

For $J_1\neq0$, the Hamiltonian in Eq.~\eqref{eq:cluster_hamiltonian} explicitly breaks the $\mathbb{Z}_2\times \mathbb{Z}_2$ symmetry generated by independent $\pi$ rotations of the even and odd sublattices about the $z$ axis:
\begin{align}
  U_{\rm e} = \prod_{j:\mathrm{even}}\sigma_j^z,\quad 
  U_{\rm o} = \prod_{j:\mathrm{odd}}\sigma_j^z.
\end{align}
We investigate the fate of this symmetry after a sudden global quench in which the system is prepared in the ground state of the Hamiltonian~\eqref{eq:cluster_hamiltonian} with $J_1 \neq 0$ and then evolved with $J_1=0$. In particular, we focus on symmetry breaking within a subsystem $A$ of length $2\ell$. This subsystem is described by the reduced density matrix
\begin{align}
\rho_A(t) = \mathrm{Tr}_{\bar{A}}[\ket{\psi_t}\!\!\bra{\psi_t}],
\end{align}
where $\ket{\psi_t}=e^{-it H(0,J_2,\delta)}\ket{\psi_0}$. In the thermodynamic limit $L\to\infty$, $\rho_A(t)$ is expected to relax at long times to a stationary state that generally respects the $\mathbb{Z}_2\times \mathbb{Z}_2$ symmetry. In other words, denoting the restriction of $U_{\rm e(o)}$ to subsystem $A$ by $U_{A,\mathrm{e(o)}}$, we expect the reduced density matrix to satisfy $[\rho_A(t\to\infty),U_{A,\mathrm{e/o}}]=0$.

To monitor the time evolution of the $\mathbb{Z}_2\times \mathbb{Z}_2$ symmetry in $A$, we employ the fidelity-based LCF defined by
\begin{align}\label{eq:LCF_cluster}
  \widetilde{L}_A^{(g)}(t)=-\log F(\rho_A(t),\rho_{A,g}(t)),\quad g\in \{\rm 1,e,o,eo\}, 
\end{align}
where $\rho_{A,g}=U_{A,g}\rho_A U_{A,g}^\dag$, with $U_{A,\mathrm{eo}}=U_{A,\mathrm{e}}U_{A,\mathrm{o}}$ and $\rho_{A,1}=\rho_A$. For comparison, we also study the EA for the same symmetry:
\begin{align}\label{eq:cluster_ea}
  A_{\mathbb{Z}_2\times \mathbb{Z}_2}(t)
  = 
  S\left(\mathcal{G}_A^{\mathbb{Z}_2\times \mathbb{Z}_2}[\rho_A(t)]\right)-S(\rho_A(t)), 
\end{align}
where $\mathcal{G}_A^{\mathbb{Z}_2 \times \mathbb{Z}_2}[\rho_A]=\sum_{g\in \{\rm 1,e,o,eo\}} \rho_{A,g}/4$.

\subsection{Time evolution of the fidelity-based LCF}
\label{subsec:cluster_gaussian}
To calculate the fidelity-based LCF, we exploit the fact that both the time-evolved state $\ket{\psi_t}$ and its reduced density matrix $\rho_A(t)$ are Gaussian because the Hamiltonian~\eqref{eq:cluster_fermion_hamiltonian} is quadratic in $(a_i,b_i)$.
In addition, the states $\rho_{A,g}=U_{A,g}\rho_AU_{A,g}^\dag$ are also Gaussian because the transformations $U_{A,g}$ ($g\in\{\rm e,o,eo\}$) preserve Gaussianity. We can therefore univocally describe $\rho_{A,g}$ by their two-point correlation matrices $\Gamma^{(g)}$~\cite{peschel2003calculation}, whose elements are given by
\begin{equation}
  \Gamma_{jj'}^{(g)}(t)
  :=
  2\mathrm{Tr}[\rho_{A,g}(t) {\bf \Psi}_j {\bf \Psi}_{j'}^\dag]
  -\delta_{j,j'}I_2,\quad j,j'\in A,
  \label{eq:cluster_correlation_matrix}
\end{equation}
where ${\bf \Psi}_j=(a_j,b_j)^\top$.
Evaluating each element and taking the thermodynamic limit $L\to\infty$, we obtain 
\begin{equation}\label{eq:cluster_toeplitz_matrix}
  \Gamma_{jj'}^{(g)}(t)
  =
  \int_{-\pi}^{\pi}\frac{dk}{2\pi}
  e^{ik(j-j')}\mathcal{T}^{(g)}(k,t),
\end{equation}
where the $2\times 2$ symbol $\mathcal{T}^{(g)}$ is given by 
\begin{align}\label{eq:toeplitz_symbol}
  \mathcal{T}^{(1)}=\mathcal{T}^{\rm (eo)}=
  \begin{pmatrix}
    \cos \theta_k & e^{-i(\phi_k+2t\varepsilon_k)}\sin\theta_k \\
    e^{i(\phi_k+2t\varepsilon_k)}\sin\theta_k & -\cos \theta_k
  \end{pmatrix},
  \quad 
  \mathcal{T}^{\rm (e)}=\mathcal{T}^{\rm (o)}= 
  \sigma^z \mathcal{T}^{(1)}\sigma^z, 
\end{align}
where $\varepsilon_{k}=\delta-J_2\cos k$ is the dispersion relation of the post-quench Hamiltonian.

Equation~\eqref{eq:toeplitz_symbol} shows that $\Gamma^{(1)}=\Gamma^{\rm (eo)}$ and $\Gamma^{\rm (e)}=\Gamma^{\rm (o)}$. The former equality implies $\rho_A=\rho_{A,{\rm eo}}$, meaning that the reduced density matrix respects the $\mathbb{Z}_2$ symmetry generated by $U_{A,{\rm eo}}$. This follows because $U_{A,{\rm eo}}=\prod_{i\in A}(1-2a_i^\dag a_i)(1-2b_i^\dag b_i)$ corresponds to the fermion-number parity operator, and any fermionic Gaussian state is invariant under the parity transformation. The latter equality implies $\rho_{A,{\rm e}}=\rho_{A,{\rm o}}$ and hence $\widetilde{L}_A^{\rm (e)}=\widetilde{L}_A^{\rm (o)}$. We therefore focus on $\widetilde{L}_A^{\rm (e)}$ below.

As shown explicitly in Appendix~\ref{app:fermionic_gaussian_fidelity}, the fidelity-based LCF in Eq.~\eqref{eq:LCF_cluster} can be expressed in terms of the two-point correlation matrices as
\footnote{This expression illustrates a technical advantage of the fidelity-based LCF: it can be computed directly from two-point correlation matrices, allowing numerically exact calculations even for large subsystems. By contrast, as pointed out by Ref.~\cite{travaglino2026gaussianasymmetrymeasure}, the group-twirled state in Eq.~\eqref{eq:cluster_ea} is generally non-Gaussian even when $\rho_A$ is Gaussian. Consequently, standard Gaussian algebra does not provide an equally direct way to compute the EA.}
\begin{align}\label{eq:LCF_trace}
  \widetilde{L}_A^{\rm (e)}(t)=-\mathrm{Tr}\log\left[
    \left(\frac{I+\Gamma^{(1)}}{2}\right)
    \left(
      I+\sqrt{\sqrt{\frac{I-\Gamma^{(1)}}{I+\Gamma^{(1)}}}\frac{I-\Gamma^{\rm (e)}}{I+\Gamma^{\rm (e)}}
      \sqrt{\frac{I-\Gamma^{(1)}}{I+\Gamma^{(1)}}}
      }
    \right)
  \right]. 
\end{align}
By expanding the right-hand side of Eq.~\eqref{eq:LCF_trace} in terms of the moments $\mathrm{Tr}[(\Gamma^{(1)})^n]$ and evaluating them using the multidimensional stationary phase method, we can exactly calculate the asymptotic form of the trace in Eq.~\eqref{eq:LCF_trace} in the hydrodynamic limit, in which $t,\ell\to\infty$ with $\zeta=t/\ell$ fixed, as 
\begin{align}\label{eq:LCF_analytic}
  \widetilde{L}_A^{\rm (e)}(t) 
  = 
  \ell \int_{-\pi}^\pi \frac{dk}{2\pi}\lambda_k\max(1-2|\varepsilon_k'|\zeta,0) +O(1). 
\end{align}
Here $\lambda_k=-\log|\cos \theta_k|$ and $\varepsilon_k'=\partial_k \varepsilon_k$ is the group velocity of the Bogoliubov quasiparticles after the quench.
We give the detailed derivation of Eq.~\eqref{eq:LCF_analytic} in Appendix~\ref{app:cluster_lcf_hydrodynamics}. 

Equation~\eqref{eq:LCF_analytic} can be understood in terms of the quasiparticle picture, which provides a semiclassical description of the dynamics following a global quench in an integrable system~\cite{Calabrese2005}.
In this picture, a global quench at $t=0$ creates entangled quasiparticle pairs uniformly throughout the system, and only the pairs that remain inside the subsystem contribute to symmetry breaking. Indeed, denoting the annihilation operators for upper- and lower-band quasiparticles with momentum $k$ after the quench by $\gamma_{k,+}$ and $\gamma_{k,-}$, respectively, $\lambda_k$ can be rewritten as
\begin{align}\label{eq:lam_k}
  \lambda_k=-\frac{1}{2}\log\left(1-4\left|\bra{\psi_0}\gamma_{k,+}^\dagger \gamma_{k,-}\ket{\psi_0}\right|^2\right).
\end{align}
Equation~\eqref{eq:lam_k} shows that the quench excites correlated pairs of quasiparticles from the upper and lower bands and that their correlations contribute to the breaking of the $\mathbb{Z}_2\times \mathbb{Z}_2$ symmetry. Since the two bands have dispersions $\pm \varepsilon_k$, the quasiparticles have opposite group velocities, $\pm \varepsilon_k'$. As a result, the two members of each pair propagate in opposite directions, and the number of pairs contained entirely within the subsystem, together with the LCF, gradually decreases over time. This effect is captured by the remaining factor in the integrand, $\max(1-2|\varepsilon_k'|\zeta,0)$, which filters out the contributions from pairs that have left the subsystem.

\begin{figure}[t]
    \raggedright
    \includegraphics[width=0.95\textwidth]{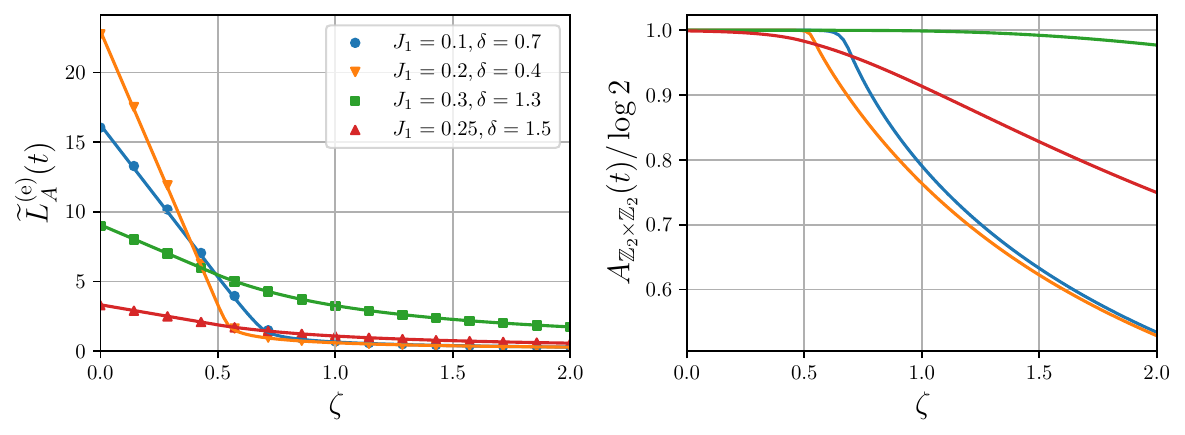}
    \caption{Time evolution of the fidelity-based LCF and the EA in the quench dynamics of the one-dimensional cluster model in Eq.~\eqref{eq:cluster_hamiltonian}. The solid lines correspond to the analytical expressions in Eqs.~\eqref{eq:LCF_analytic} and \eqref{eq:cluster_ea_hydrodynamic} for the hydrodynamic limit $\ell,t\gg1$. The dotted lines are the exact result obtained by numerically evaluating Eq.~\eqref{eq:LCF_trace}. We set $J_2=1$ and $\ell=128$ for all the plots.} 
    \label{fig:lcf_and_ea_cluster}
\end{figure}
In Fig.~\ref{fig:lcf_and_ea_cluster}, we plot the LCF as a function of time for several initial configurations.
The figure shows that the analytic prediction in Eq.~\eqref{eq:LCF_analytic} agrees well with the exact result obtained by numerically evaluating Eq.~\eqref{eq:LCF_trace}.
In addition, some pairs of the curves corresponding to different initial states intersect at certain times, indicating the occurrence of the quantum Mpemba effect.

The quasiparticle picture gives an exact condition and a clear interpretation for the quantum Mpemba effect in the hydrodynamic limit. Consider two systems prepared in different initial states labeled by I and II, and let $\lambda_k^{(\mathrm{I})}$ and $\lambda_k^{(\mathrm{II})}$ denote their respective momentum-resolved weights in the LCF. Then, the quantum Mpemba effect occurs when the following two conditions are met: (i) system $\mathrm{I}$ initially exhibits stronger symmetry breaking than system $\mathrm{II}$ and (ii) system $\mathrm{II}$ exhibits stronger symmetry breaking than system $\mathrm{I}$ for $\zeta>\zeta_*$, where $\zeta_*$ denotes the rescaled time after which the ordering is reversed. On one hand, according to Eq.~\eqref{eq:LCF_analytic}, condition (i) is equivalent to
\begin{align}
  \int_{-\pi}^{\pi}\frac{dk}{2\pi}\lambda_k^{(\mathrm{I})}
  &>
  \int_{-\pi}^{\pi}\frac{dk}{2\pi}\lambda_k^{(\mathrm{II})},
  \label{eq:cluster_mpemba_initial_condition}
\end{align}
which means that the total pair correlation of quasiparticles in the initial state of system $\mathrm{I}$ makes a larger total contribution to symmetry breaking. On the other hand, condition (ii) is equivalent to
\begin{align}
  \int_{|\varepsilon_k'|<\frac{1}{2\zeta}}\frac{dk}{2\pi}\lambda_k^{(\mathrm{I})}(1-2|\varepsilon_k'|\zeta)
  &<
  \int_{|\varepsilon_k'|<\frac{1}{2\zeta}}\frac{dk}{2\pi}
  \lambda_k^{(\mathrm{II})}(1-2|\varepsilon_k'|\zeta), 
  \qquad \zeta>\zeta_*.
  \label{eq:cluster_mpemba_late_condition}
\end{align}
This inequality means that the pair correlation of slow quasiparticle pairs remaining inside the subsystem at large times is larger in system $\mathrm{II}$. Therefore, the quantum Mpemba effect occurs when system $\mathrm{I}$ has a larger total pair correlation of quasiparticles but a smaller pair correlation of slow quasiparticles. A closely related argument based on the quasiparticle picture has been identified for quantum Mpemba effects associated with $U(1)$ symmetry and Cooper pairs~\cite{Murciano2023,Yamashika2025}; the present LCF analysis extends this picture to a finite-group symmetry.

\subsection{Time evolution of EA}

For comparison with the fidelity-based LCF, we calculate the EA in Eq.~\eqref{eq:cluster_ea} in the hydrodynamic limit. To this end, we introduce the R\'enyi EA
\begin{align}
  A_{\mathbb{Z}_2\times \mathbb{Z}_2}^{(n)}(t)
  =
  S_n\left(\mathcal{G}_A^{\mathbb{Z}_2\times \mathbb{Z}_2}[\rho_A(t)]\right)
  -S_n(\rho_A(t)),
  \label{eq:cluster_renyi_ea}
\end{align}
where $S_n(\rho)=(1-n)^{-1}\log\mathrm{Tr}\rho^n$ is the R\'enyi entropy. The EA can be obtained by taking the von Neumann limit in Eq.~\eqref{eq:cluster_renyi_ea} as $\lim_{n\to1}A_{\mathbb{Z}_2\times \mathbb{Z}_2}^{(n)}(t)=A_{\mathbb{Z}_2\times \mathbb{Z}_2}(t)$. Using the identities $\rho_{A,1}=\rho_{A,{\rm eo}}$ and $\rho_{A,{\rm e}}=\rho_{A,{\rm o}}$, the R\'enyi EA can be written as 
\begin{align}\label{eq:REA}
  A_{\mathbb{Z}_2\times \mathbb{Z}_2}^{(n)} = \frac{1}{1-n}\log
  \left(\frac{1}{2^n}\sum_{\boldsymbol{g}\in\{1,{\rm e}\}^{\otimes n}} \frac{Z_n(\boldsymbol{g},t)}{Z_n(\boldsymbol{1},t)}\right),
\end{align}
where 
\begin{align}
  Z_n(\boldsymbol{g},t)= \mathrm{Tr}\left[\prod_{j=1}^n \rho_{A,g_j}(t)\right]
\end{align}
are the charged moments. 

The charged moments can be expressed in terms of the two-point correlation matrices as~\cite{hara2026}
\begin{align}\label{eq:charged_moment_Gamma}
  Z_n(\boldsymbol{g},t)
  = 
  \det\left[
    \left(\prod_{j=1}^n \frac{I+\Gamma^{(g_j)}(t)}{2}\right)
    \left(I+\prod_{j=1}^n \frac{I-\Gamma^{(g_j)}(t)}{I+\Gamma^{(g_j)}(t)} \right)
  \right]. 
\end{align}
As in the derivation of Eq.~\eqref{eq:LCF_analytic}, the asymptotic form of the charged moments in the hydrodynamic limit can be derived by expanding Eq.~\eqref{eq:charged_moment_Gamma} in terms of the moments of the correlation matrices, $\mathrm{Tr}[(\Gamma^{(1)})^m]$, and evaluating it using the multidimensional stationary phase method. It results in 
\begin{align}\label{eq:CM_asymptotic}
  \frac{Z_n(\boldsymbol{g},t)}{Z_n(\boldsymbol{1},t)} 
  \simeq X_\zeta^{N_{\rm dw}(\boldsymbol{g})} ,
\end{align}
with 
\begin{gather}
  X_\zeta=\exp\left(-\ell \int_{-\pi}^\pi \frac{dk}{2\pi}\lambda_k\max(1-2\zeta|\varepsilon_k'|,0)\right),
  \label{eq:cluster_X_zeta}\\
  N_{\rm dw}(\boldsymbol{g}) = \sum_{j=1}^n (1-\delta_{g_j,g_{j+1}}),\quad g_{n+1}=g_1. 
\end{gather}
Substituting Eq.~\eqref{eq:CM_asymptotic} into Eq.~\eqref{eq:REA} and taking the summation over $\boldsymbol{g}\in\{1,\mathrm{e}\}^{\otimes n}$, we obtain 
\begin{align}
  A_{\mathbb{Z}_2\times \mathbb{Z}_2}^{(n)}(t)
  \simeq 
  \frac{1}{1-n}\log\left( \left[\frac{1+X_\zeta}{2}\right]^n+\left[\frac{1-X_\zeta}{2}\right]^n\right). 
\end{align}
Taking the von Neumann limit $n\to1$, it reduces to 
\begin{align}\label{eq:cluster_ea_hydrodynamic}
  A_{\mathbb{Z}_2\times \mathbb{Z}_2}(t)
  \simeq 
  -\frac{1+X_\zeta}{2}\log \frac{1+X_\zeta}{2}
  -\frac{1-X_\zeta}{2}\log \frac{1-X_\zeta}{2}. 
\end{align}

Equation~\eqref{eq:cluster_ea_hydrodynamic} directly relates the EA to the fidelity-based LCF. Since $X_\zeta\simeq \exp(-\widetilde{L}_A^{\rm(e)})$ from Eqs.~\eqref{eq:LCF_analytic} and \eqref{eq:cluster_X_zeta}, the EA in Eq.~\eqref{eq:cluster_ea_hydrodynamic} decreases as the fidelity-based LCF decreases and vanishes when the LCF vanishes. In this sense, the two quantities exhibit the same qualitative behavior, as expected for faithful asymmetry monotones. However, for large $\ell$, the EA saturates at $\log2$ since $X_\zeta$ in Eq.~\eqref{eq:cluster_X_zeta} vanishes in that limit. As discussed above, this saturation reflects the fact that the EA, unlike the fidelity-based LCF, is not extensive. \footnote{Although the symmetry group $\mathbb{Z}_2\times\mathbb{Z}_2$ has order $4$, the identities $\rho_{A,\mathrm{e}}=\rho_{A,\mathrm{o}}$ and $\rho_A=\rho_{A,\mathrm{eo}}$ due to the Gaussianity of the states reduce the four symmetry images to two distinct states, giving the saturation value $\log2$ rather than $\log |\mathbb{Z}_2\times \mathbb{Z}_2|=2\log2$.}

The extensivity is crucial for diagnosing the quantum Mpemba effect of the finite group symmetry. Figure~\ref{fig:lcf_and_ea_cluster} shows the EA as a function of time for several initial states. Because the EA is not extensive, at $t=0$ it approaches the same value, $\log2$, for all initial states as $\ell$ increases, independently of how strongly each state breaks the symmetry. The initial states cannot then be ordered by their degree of symmetry breaking, which prevents the quantum Mpemba effect from being identified through the EA, as already pointed out in Ref.~\cite{Ferro2023}. By contrast, the fidelity-based LCF remains extensive, and its density retains the dependence on the initial state. It therefore distinguishes how strongly each initial state breaks the symmetry and allows the quantum Mpemba effect to be discussed unambiguously as described in Sec.~\ref{sec:cluster_mpemba}.

\section{LCF Ordering Reversal Missed by Entanglement Asymmetry in the XXX Chain}
\label{sec:su2_mps}

The XXZ benchmark showed that the LCF components can refine a crossing already visible in EA.
We now demonstrate a stronger advantage in an interacting many-body system: one LCF component reverses the ordering of two initial states even though EA retains the same ordering throughout the studied time window.
This provides a direct many-body example in which compressing the dependence on the group element into a single scalar discards dynamical information retained by the LCF family.
Although the XXX Hamiltonian is $SU(2)$ invariant, we evaluate both diagnostics with respect to the quaternion subgroup $Q_8\subset SU(2)$.

\subsection{Setup and numerical results}

We consider the open spin-$\frac12$ XXX chain
\begin{equation}
  H_{\rm XXX}
  =
  \frac14\sum_{j=1}^{N-1}
  \bm{\sigma}_j\!\cdot\!\bm{\sigma}_{j+1},
\end{equation}
with the exchange coupling set to unity.
For the initial state, we take the two-sublattice product family
\begin{equation}
  |\Psi_{\theta_A,\theta_B;\beta}(0)\rangle
  :=
  \bigotimes_{j\,{\rm odd}}
  \ex{-i\frac{\beta}{2}\sigma_j^y}|\theta_A\rangle_j
  \otimes
  \bigotimes_{j\,{\rm even}}
  \ex{-i\frac{\beta}{2}\sigma_j^y}|\theta_B\rangle_j,
\end{equation}
where
\begin{equation}
  |\theta\rangle
  :=
  \cos\frac{\theta}{2}|\uparrow\rangle
  +
  \sin\frac{\theta}{2}|\downarrow\rangle.
\end{equation}
We consider the following two initial states:
\begin{align}
  |\Psi_{\rm H}(0)\rangle
  &=
  |\Psi_{\fr{\pi}{4},\,\fr{3\pi}{4};\,\fr{\pi}{3}}(0)\rangle,
  \\
  |\Psi_{\rm C}(0)\rangle
  &=
  |\Psi_{\fr{\pi}{6},\,\fr{3\pi}{4};\,0}(0)\rangle.
\end{align}
We refer to them as State~H and State~C.
We choose the subsystem $A$ to be the central block of length $\ell=4$ and define
\begin{equation}
  \rho_{A,s}(t)
  :=
  \tr_{\bar A}\Bigl(
    \ex{-itH_{\rm XXX}}
    |\Psi_s(0)\rangle\!\langle\Psi_s(0)|
    \ex{itH_{\rm XXX}}
  \Bigr),
  \qquad
  s={\rm H,C}.
\end{equation}

To compare EA with the individual LCF components, we restrict the symmetry diagnostics to the quaternion subgroup
\begin{equation}
  Q_8
  =
  \{\pm \mathbf{1},\,\pm i\sigma^x,\,\pm i\sigma^y,\,\pm i\sigma^z\}
  \subset SU(2)
\end{equation}
in the spin-$\frac12$ representation.
Let $g_x,g_y,g_z\in Q_8$ denote the elements represented by $-i\sigma^x$, $-i\sigma^y$, and $-i\sigma^z$.
On the subsystem $A$, they act as
\begin{equation}
  U_A(g_\mu)
  =
  \ex{-i\pi S_A^\mu},
  \qquad
  S_A^\mu
  =
  \frac12\sum_{j\in A}\sigma_j^\mu,
  \qquad
  \mu=x,y,z.
\end{equation}
Since $g$ and $-g$ induce the same conjugation action on density matrices, the $Q_8$ twirling map reduces to
\begin{align}
  \mathcal{G}_A^{Q_8}[\rho]
  :=&
  \frac18\sum_{g\in Q_8}U_A(g)\rho U_A(g)^\dagger
  \\
  =&
  \frac14\Bigl(
    \rho
    +U_A(g_x)\rho U_A(g_x)^\dagger
    +U_A(g_y)\rho U_A(g_y)^\dagger
    +U_A(g_z)\rho U_A(g_z)^\dagger
  \Bigr).
\end{align}
As the incomplete benchmark, we use
\begin{equation}
  A_{Q_8}(\rho_{A,s}(t))
  :=
  S\!\bigl(\mathcal{G}_A^{Q_8}[\rho_{A,s}(t)]\bigr)
  -
  S(\rho_{A,s}(t)).
\end{equation}

We compare this EA with the fidelity-based LCF components
\begin{equation}
  \widetilde{L}_{A,s}^{(\mu)}(t)
  :=
  -\log F\!\Bigl(
    \rho_{A,s}(t),
    U_A(g_\mu)\rho_{A,s}(t)U_A(g_\mu)^\dagger
  \Bigr).
\end{equation}
We focus on the $g_x$ and $g_z$ components, which already exhibit the contrast central to this section.
We use exact diagonalization for $N=8$ and TEBD for $N=16,24,32,40$.

\begin{figure}[t]
  \centering
  \includegraphics[width=\textwidth]{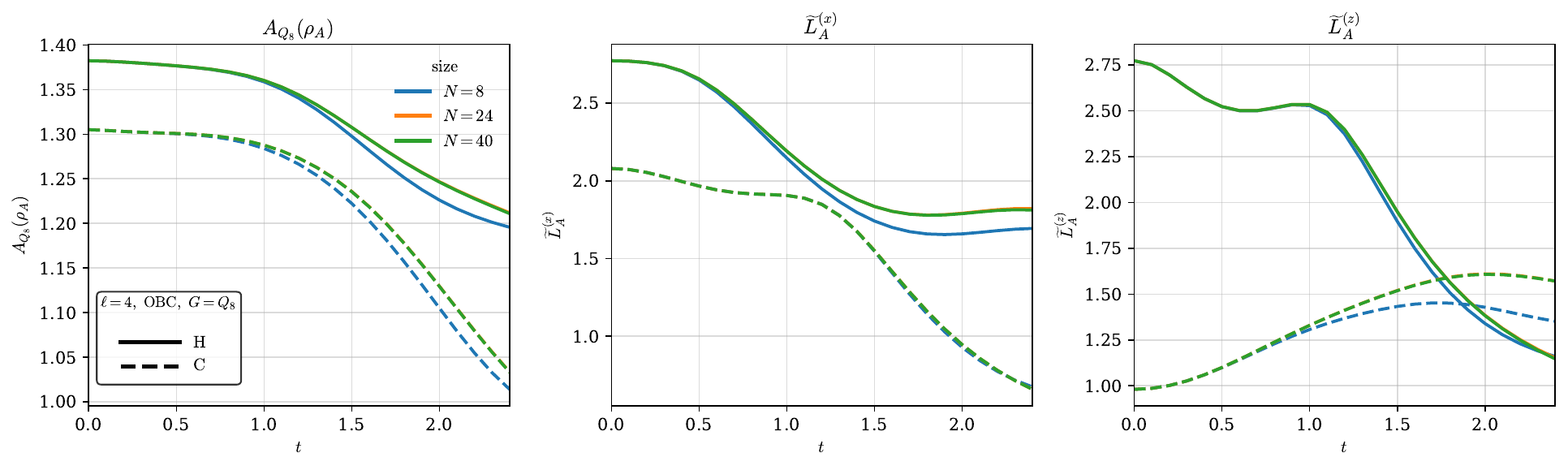}
  \caption{
  Entanglement asymmetry and the $g_x$ and $g_z$ components of the fidelity-based LCF in the open XXX chain.
  The result for $N=8$ is obtained by exact diagonalization. 
  The results for $N=24$ and $40$ are obtained by the TEBD algorithm. The TEBD algorithm is implemented by using the second-order Suzuki-Trotter decomposition with a time step $\delta t=0.1$ and setting the bond dimension $\chi=24$. The solid and dashed lines show the results for State~H and State~C, respectively.
  The left, middle, and right panels show $A_{Q_8}$, $\widetilde{L}_A^{(x)}$, and $\widetilde{L}_A^{(z)}$, respectively. The EA and the $g_x$ LCF component retain their initial ordering throughout the plotted time window, whereas the $g_z$ component exhibits a crossing for every displayed size. We confirmed the same qualitative behavior for $N=16,32$. }
  \label{fig:su2_mps_largeN}
\end{figure}

Fig.~\ref{fig:su2_mps_largeN} reveals a sharp contrast between EA and the individual LCF components.
Neither $A_{Q_8}$ nor $\widetilde{L}_A^{(x)}$ crosses within the interval $0\le t\le2.4$.
By contrast, $\widetilde{L}_A^{(z)}$ crosses once for every system size $N=8,16,24,32,40$ studied.
For $N\ge16$, the crossing time is already concentrated near
\begin{equation}
  t_z^\ast\simeq1.78.
\end{equation}
The weak size dependence shows that the ordering reversal is robust against increasing the total system size and is not an artifact of the smallest chain.
Most importantly, this reversal is completely absent from EA.

Because the State~C curve of $\widetilde{L}_A^{(z)}(t)$ is non-monotonic, we refer to this phenomenon as an ordering reversal rather than, by itself, a conventional Mpemba effect.
This distinction does not weaken the central conclusion: EA preserves the initial ordering, while the $g_z$ LCF component reverses it.
The LCF family therefore retains dynamical information that is lost in the EA.

The XXZ and XXX results demonstrate two complementary advantages of retaining the LCF family.
In the XXZ chain, EA and both nontrivial LCF components exhibit crossings, generally at different times, so the LCF refines a Mpemba effect already visible in EA.
In the XXX chain, EA does not cross, whereas the $g_z$ component does, so the LCF reveals an ordering reversal that EA misses completely.
Together, these examples show that the LCF family can both distinguish the internal structure of a scalar crossing and retain a crossing that is absent after group averaging.

\section{Quantum Mpemba Effect in a Single Qubit}
\label{sec:depolarizing_qubit}

In the previous section, we found that the ordering of two states according to an individual LCF can reverse even when their ordering according to the EA does not. We here show that, for a single qubit, this difference can be analyzed more explicitly because the EA and the LCFs have closed form expressions. We first derive the exact relation between the EA and LCFs, clarifying how it allows an individual LCF to capture a quantum Mpemba effect that is absent from the EA. We then demonstrate this possibility in the dynamics of a qubit under depolarizing noise. 

\subsection{Exact relation between the EA and the LCF Components}
\label{subsec:qubit_mpemba}

An arbitrary qubit state can be written in terms of the Bloch vector $\bm r=(r_x,r_y,r_z)^\top$ as
\begin{equation}
  \rho
  =\frac12\left(
    I+\sum_{\mu=x,y,z}r_\mu\sigma^\mu
  \right).
  \label{eq:qubit_bloch_state}
\end{equation}
We consider the $\mathbb{Z}_2\times \mathbb{Z}_2$ action on density matrices generated by conjugation with $U_x=\sigma^x$ and $U_z=\sigma^z$. Up to an irrelevant phase, the third nonidentity transformation is represented by $U_y=\sigma^y$. For each $\mu=x,y,z$, the fidelity-based LCF associated with $U_\mu$ is
\begin{equation}
  \widetilde{L}^{(\mu)}_\rho
  =
  -\log F\left(\rho,U_\mu\rho U_\mu^\dagger\right),
  \qquad \mu=x,y,z.
  \label{eq:qubit_lcf_definition}
\end{equation}
To evaluate Eq.~\eqref{eq:qubit_lcf_definition}, we use the Uhlmann fidelity for two qubit states with Bloch vectors $\bm r$ and $\bm r'$:
\begin{equation}\label{eq:qubit_fidelity}
  F(\rho,\rho')
  =
  \sqrt{
    \frac{
      1+\bm r\cdot\bm r'
      +\sqrt{(1-|\bm r|^2)(1-|\bm r'|^2)}
    }{2}
  }.
\end{equation}
Conjugation by $\sigma^\mu$ leaves the component $r_\mu$ of the Bloch vector unchanged and reverses the signs of the other two components. Substituting the resulting Bloch vector of $U_\mu\rho U_\mu^\dagger$ into Eq.~\eqref{eq:qubit_fidelity}, we obtain
\begin{equation}
  \widetilde{L}^{(\mu)}_\rho
  =-\frac12\log\!\left(1-|\bm r|^2+r_\mu^2\right),
  \qquad \mu=x,y,z.
  \label{eq:qubit_lcf_components}
\end{equation}
Equation~\eqref{eq:qubit_lcf_components} shows that $\widetilde{L}^{(\mu)}_\rho$ depends on the two components of the Bloch vector whose signs are reversed by $U_\mu$. Taken together, the three LCFs therefore retain information about each component of the Bloch vector.

On the other hand, the EA for the present $\bb{Z}_2\times\bb{Z}_2$ symmetry is given by
\begin{align}
  A_{\bb{Z}_2\times\bb{Z}_2}(\rho)
  =\log 2
  +\frac{1+|\boldsymbol{r}|}{2}\log\frac{1+|\boldsymbol{r}|}{2}
  +\frac{1-|\boldsymbol{r}|}{2}\log\frac{1-|\boldsymbol{r}|}{2}.
  \label{eq:qubit_ea}
\end{align}
Equation~\eqref{eq:qubit_ea} shows that the EA is determined only by the length of the Bloch vector and, in this sense, contains less information than the full LCF family. Indeed, exponentiating Eq.~\eqref{eq:qubit_lcf_components} and summing over $\mu=x,y,z$ gives
\begin{equation}
  |\bm r|^2
  =\frac12\sum_{\mu=x,y,z}
  \left(1-e^{-2\widetilde{L}^{(\mu)}_\rho}\right).
  \label{eq:qubit_ea_from_lcf}
\end{equation}
Equations~\eqref{eq:qubit_ea} and \eqref{eq:qubit_ea_from_lcf} show that the full LCF family determines the EA. Conversely, the EA cannot determine the three LCFs because it contains no information about the direction of the Bloch vector.

This loss of information about the direction of the Bloch vector can be crucial for the quantum Mpemba effect: during symmetry-restoration dynamics, one LCF can detect a quantum Mpemba effect that is absent in the EA. To see this, consider two systems with states $\rho_{\mathrm I}(t)$ and $\rho_{\mathrm{II}}(t)$ at time $t$. Combining Eqs.~\eqref{eq:qubit_ea} and \eqref{eq:qubit_ea_from_lcf}, and using the fact that the EA is strictly increasing with the Bloch-vector length, we obtain 
\begin{align}
  \mathrm{sgn}\left[
    A_{\bb{Z}_2\times\bb{Z}_2}\!\left(\rho_{\mathrm I}(t)\right)
    -A_{\bb{Z}_2\times\bb{Z}_2}\!\left(\rho_{\mathrm{II}}(t)\right)
  \right]
  =
  \mathrm{sgn}\!\left[
    \sum_{\mu=x,y,z}
    \left(
      e^{-2\widetilde{L}^{(\mu)}_{\rho_{\mathrm{II}}(t)}}
      -e^{-2\widetilde{L}^{(\mu)}_{\rho_{\mathrm I}(t)}}
    \right)
  \right].
  \label{eq:qubit_ea_ordering}
\end{align}
Equation~\eqref{eq:qubit_ea_ordering} directly relates the criterion for a quantum Mpemba effect in the EA to the corresponding criteria for the three LCFs. If the EA is initially larger for system $\mathrm I$ than for system $\mathrm{II}$, it exhibits the effect when the sign on the left-hand side changes from positive to negative. Likewise, if $\widetilde{L}^{(\mu)}$ is initially larger for system $\mathrm I$, it exhibits the quantum Mpemba effect when the sign of the corresponding term inside the sum on the right-hand side changes from positive to negative. A sign change in one term need not change the sign of the sum, which allows an individual LCF to exhibit the quantum Mpemba effect while the EA does not.

For symmetry-restoration dynamics, a quantum Mpemba effect in the EA necessarily implies a quantum Mpemba effect in at least one of the three LCFs if all three LCFs are initially larger for system $\mathrm I$. Indeed, all three terms on the right-hand side of Eq.~\eqref{eq:qubit_ea_ordering} are then positive at $t=0$, and the EA is also larger for system $\mathrm I$. If the EA becomes smaller for system $\mathrm I$ at a later time, the sum must become negative, and at least one term indexed by some $\mu$ must have changed sign. For that $\mu$, the LCF has changed from being larger for system $\mathrm I$ to being larger for system $\mathrm{II}$, indicating the occurrence of the quantum Mpemba effect. 

\subsection{Depolarizing noise}
\label{subsec:qubit_dynamics}

We now demonstrate that an individual LCF can exhibit the quantum Mpemba effect while the EA does not in the dynamics of a qubit under depolarizing noise, which is a standard model of qubit decoherence. In this model, random Pauli errors progressively erase the information encoded in the Bloch vector and drive the state toward the maximally mixed state. The time evolution is described by the master equation
\begin{equation}
  \frac{\partial \rho(t)}{\partial t}
  =\frac12\sum_{\mu=x,y,z}
  \kappa_\mu\left(\sigma^\mu\rho(t)\sigma^\mu-\rho(t)\right).
  \label{eq:qubit_lindblad}
\end{equation}
Here $\kappa_\mu>0$ is the rate of the Pauli error channel generated by $\sigma^\mu$. The Lindbladian is covariant under conjugation by $U_x$, $U_y$, and $U_z$.
If we denote as $r_\mu(t)$ ($\mu=x,y,z$) each element of the Bloch vector of $\rho(t)$, Eq.~\eqref{eq:qubit_lindblad} can be solved as 
\begin{equation}
  r_\mu(t)=r_\mu(0)e^{-\gamma_\mu t},
  \label{eq:qubit_bloch_evolution}
\end{equation}
where $\gamma_x=\kappa_y+\kappa_z$, $\gamma_y=\kappa_x+\kappa_z$, and $\gamma_z=\kappa_x+\kappa_y$. At large times, every component of the Bloch vector vanishes and the state approaches the maximally mixed state, restoring the $\mathbb{Z}_2\times \mathbb{Z}_2$ symmetry generated by $U_x$ and $U_z$.

\begin{figure}[t]
    \raggedright
    \includegraphics[width=0.95\linewidth]{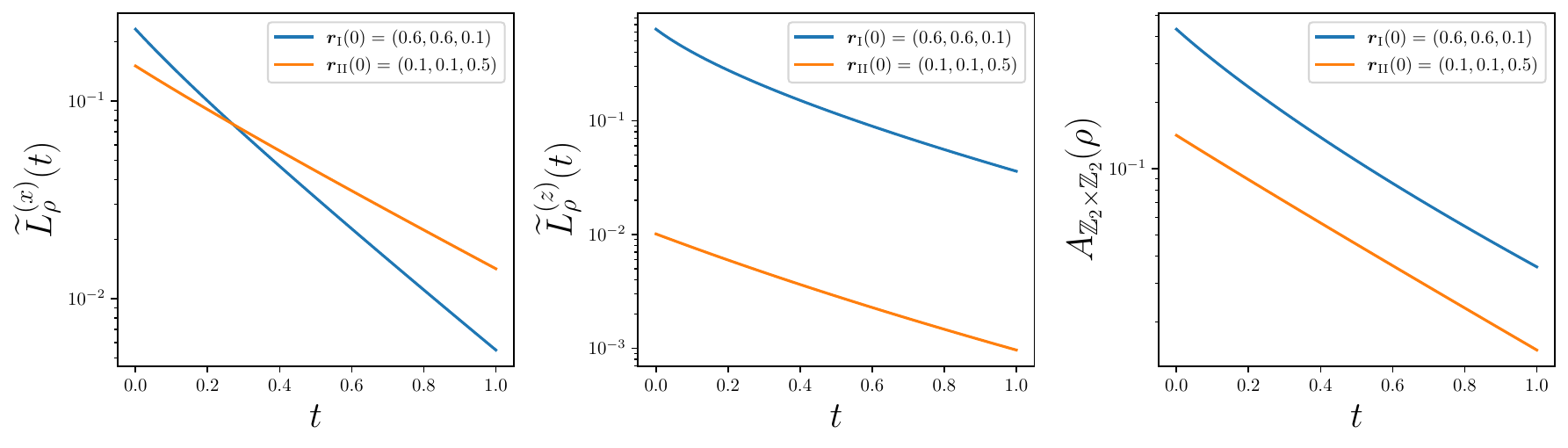}
    \caption{Time evolution of the fidelity-based LCF and the EA for the single qubits under the depolarizing noise in Eq.~\eqref{eq:qubit_lindblad}. The initial states are respectively given by Bloch vectors $\boldsymbol{r}_{\rm I}(0)$ and $\boldsymbol{r}_{\rm II}(0)$.  We set $(\kappa_x,\kappa_y,\kappa_z)=(1,0.1,0.8)$ for all the plots.
    }
    \label{fig:depolarizing}
\end{figure}
Figure~\ref{fig:depolarizing} shows the time evolution of the LCF and the EA for two initial states. We obtain these results by substituting Eq.~\eqref{eq:qubit_bloch_evolution} into Eqs.~\eqref{eq:qubit_lcf_components} and \eqref{eq:qubit_ea}. In this figure, one finds that the curves of the LCF associated with $U_x$ intersect at certain times, while neither the LCF of $U_z$ nor the EA shows a crossing. This clearly demonstrates that the LCF family captures a quantum Mpemba effect missed by the EA. 

\section{Conclusion}
\label{sec:conclusion}

We have established the LCF family as a diagnostic of discrete-symmetry breaking and restoration,
with the string order parameter as its pure-state realization and the fidelity-based LCF as its natural mixed-state extension.
The principal conceptual lesson is that faithfulness alone does not guarantee that a symmetry diagnostic retains the information needed to characterize many-body dynamics.
EA is a valid faithful monotone that compares a state with one group-twirled state,
whereas the LCF family retains one component for each group element.
This distinction between a single scalar and the full LCF family unifies the two kinds of information loss demonstrated in this work.

The XXZ analysis first verifies that the fidelity-based LCF reproduces established many-body Mpemba crossings detected by EA.
The quench dynamics of the cluster model then demonstrates a qualitative thermodynamic advantage.
For a finite group, EA has vanishing density in the thermodynamic limit,
whereas the LCF can retain an extensive contribution with a state- and time-dependent coefficient and thereby resolve a discrete-symmetry Mpemba effect even when EA approaches the same saturated value for different initial states.
Independently of this thermodynamic distinction, the XXX and qubit examples show that averaging over the group in EA can hide ordering reversals in individual LCF components.
The XXX result should be interpreted as an ordering reversal in one LCF component rather than, by itself, as a conventional Mpemba effect,
while the analytically solvable qubit model provides a setting in which an LCF component displays a Mpemba crossing that is absent in EA.

For pure states and finite groups, the operational importance of retaining the full family is formalized by the completeness of the LCF for exact i.i.d.\ conversion.
We have not assumed or established an analogous completeness theorem for arbitrary mixed states.
Instead, our results demonstrate that retaining the full LCF family remains physically useful for mixed reduced states,
where individual components can retain thermodynamic and dynamical information that is discarded in the EA.
The replica construction and RCFT analysis further show that this information is analytically accessible.
Together, these results provide a concrete many-body rationale for taking completeness, and more generally information preservation, seriously when choosing diagnostics of symmetry breaking.

A natural next step is to determine which mixed-state conversion tasks, if any, are completely characterized by the fidelity-based LCF family.
It is also important to establish general conditions under which each LCF component has a well-defined density in the large-subsystem limit and these densities govern the ordering and crossings of symmetry-restoration curves.

\section*{Acknowledgments}
We thank Sara Murciano for discussions.
YK is supported by the INAMORI Frontier Program at Kyushu University and JSPS KAKENHI Grant Number 23K20046.
HT is supported by JSPS Grants-in-Aid for Scientific Research No. JP25K00924, MEXT KAKENHI Grant-in-Aid for Transformative
Research Areas B ``Quantum Energy Innovation'' Grant Numbers 24H00830 and 24H00831, JST FOREST No. JPMJFR2365, JST MOONSHOT No. JPMJMS256E and Royal Society International Collaboration Awards 2025 Flexigrant number ICA/R2/252240.
SY acknowledges support from JSPS KAKENHI Grants No.~JP25K23355 and JP26K17050, and from the Institute for Advanced Science, University of Electro-Communications.

\appendix

\section{Derivation of the CFT Scaling}
\label{app:rcft_lcf}

We derive Eq.~\eqref{eq:CFT_LCF_scaling} by evaluating the replica geometry in the open channel of the strip.
The symmetry defect $\ca{L}_a$ can be moved to the Euclidean boundary,
where it changes the boundary condition from $b$ to $ab$.
The relevant open-channel Hilbert space therefore consists of states compatible with the boundary conditions $b$ and $a b$.
In the long-interval limit $\ell\gg\beta$,
the leading contribution is determined by the state of the smallest conformal weight in this Hilbert space.

\subsection{Boundary spectrum}

For the Cardy states defined in Eq.~\eqref{eq:Cardy_state_def},
the Hilbert space on an interval with boundary conditions $p$ and $q$ is
\begin{equation}
  \mathcal{H}_{p,q}
  =
  \bigoplus_{i\in\mathcal{I}}
  N_{i p}^{\,q}\,\mathcal{V}_i,
\end{equation}
where $\mathcal{V}_i$ is the conformal family built on the primary of weight $h_i$.
The fusion coefficient $N_{i p}^{\,q}$ therefore determines which boundary excitations can propagate between $p$ and $q$.
For the symmetry action
\begin{equation}
  \mathcal{L}_a\ket{B_b}=\ket{B_{ab}},
\end{equation}
the lowest conformal weight in the channel between $b$ and $ab$ is
\begin{equation}
  h_{\mathrm{min}}
  =
  \min\Bigl\{
  h_i
  \ \Big|\ 
  N_{i b}^{\,ab}\neq0
  \Bigr\}.
\end{equation}
If $ab=b$, the identity is present and $h_{\mathrm{min}}=0$.
If $ab\neq b$, the identity is absent and unitarity gives $h_{\mathrm{min}}>0$.

The state \eqref{eq:phib} is represented by a Euclidean strip of width $\beta/2$.
When this strip is viewed as propagation along the spatial direction,
a state of conformal weight $h$ contributes
\begin{equation}
  \exp\!\left[-\frac{2\pi\ell}{\beta}h\right]
\end{equation}
over a distance $\ell$,
up to factors that do not grow with $\ell$.
This fixes the factor $2\pi/\beta$ in Eq.~\eqref{eq:CFT_LCF_scaling}.

\subsection{Replica geometry}

The reduced density matrix $\rho_A^{(b)}$ is represented by the same strip with a cut along $A$.
The action of $U_A(a)$ inserts the symmetry line along the cut.
Because the line is topological in the bulk,
it can be moved to the Euclidean boundary,
where it changes the boundary condition from $b$ to $ab$.

For a positive integer $n$, the replica moment introduced in Section~\ref{sec:replica_lcf} is
\begin{equation}
  \mathcal{Z}_n(a)
  =
  \tr\!\left[
  \bigl(\rho_A^{(b)}\sigma_A^{(b)}(a)\bigr)^n
  \right].
\end{equation}
In the channel describing propagation along the interval,
the replicated geometry contains $n$ channels from $b$ to $ab$ and $n$ channels from $ab$ back to $b$.
Its lowest conformal weight is therefore
\begin{equation}
  2n h_{\mathrm{min}}.
\end{equation}
The long-interval behavior is consequently
\begin{equation}
  \label{eq:CFT_replica_asymptotic}
  \log\frac{\mathcal{Z}_n(a)}{\mathcal{Z}_n(e)}
  =
  -\frac{2\pi\ell}{\beta}2n h_{\mathrm{min}}+O(1),
  \qquad
  \ell\gg\beta.
\end{equation}
Taking the ratio removes the Casimir energy
and the normalization factors common to the two replica geometries.

The analytic continuation to $n=1/2$ gives
\begin{equation}
  \mathcal{Z}_{1/2}(a)
  =
  F\!\left(\rho_A^{(b)},\sigma_A^{(b)}(a)\right),
  \qquad
  \mathcal{Z}_{1/2}(e)
  =
  \tr\rho_A^{(b)}
  =
  1.
\end{equation}
Equation~\eqref{eq:CFT_replica_asymptotic} then yields
\begin{equation}
  -\log F\!\left(\rho_A^{(b)},\sigma_A^{(b)}(a)\right)
  =
  \frac{2\pi\ell}{\beta}h_{\mathrm{min}}+O(1),
  \qquad
  ab\neq b.
\end{equation}
When $ab=b$,
the two reduced density matrices are identical and the LCF vanishes exactly.

\subsection{Ising CFT}

The Ising CFT has Cardy boundary conditions labeled by $\mathbf{1}$, $\epsilon$, and $\sigma$,
which correspond to the two fixed boundary conditions and the free boundary condition.
The nontrivial $\bb{Z}_2$ symmetry is represented by $\mathcal{L}_\epsilon$,
and the relevant fusion rules are
\begin{equation}
  \epsilon\times\mathbf{1}=\epsilon,
  \qquad
  \epsilon\times\epsilon=\mathbf{1},
  \qquad
  \epsilon\times\sigma=\sigma.
\end{equation}
The symmetry therefore exchanges the two fixed boundary conditions and leaves the free boundary condition unchanged.
The channel between the two fixed boundary conditions begins with the $\epsilon$ family,
whose lowest conformal weight is
\begin{equation}
  h_\epsilon=\frac12.
\end{equation}
It follows that
\begin{equation}
  \widetilde{L}_A^{(\mathbf{1})}(\epsilon)
  =
  \widetilde{L}_A^{(\epsilon)}(\epsilon)
  =
  \frac{\pi\ell}{\beta}+O(1),
  \qquad
  \widetilde{L}_A^{(\sigma)}(\epsilon)=0.
\end{equation}

\section{Strong and Weak Symmetry for Mixed States}
\label{app:strong}

For mixed states, two inequivalent notions of symmetry are commonly distinguished \cite{Groot2021,Lessa2024,Sala2024,Kusuki2026}.
This distinction explains why the fidelity-based LCF and the trace-based LCF discussed in Section~\ref{sec:lcf_resource_theory} behave differently.

Throughout the main text, a state is called symmetric when it is invariant under conjugation,
\begin{equation}
  \label{eq:weak_symmetry}
  U(g)\rho U(g)^\dagger=\rho,
  \qquad
  \forall g\in G.
\end{equation}
This condition, conventionally called \emph{weak symmetry},
is precisely the definition of the free states in Eq.~\eqref{eq:symmetric_states}.

A state has \emph{strong symmetry} if
\begin{equation}
  \label{eq:strong_symmetry}
  U(g)\rho=\mathrm{e}^{i\theta_g}\rho,
  \qquad
  \forall g\in G,
\end{equation}
where $\mathrm{e}^{i\theta_g}$ is a one-dimensional representation of $G$.
For an Abelian symmetry,
this means that the support of $\rho$ lies within a single charge sector.
Equation~\eqref{eq:strong_symmetry} immediately implies Eq.~\eqref{eq:weak_symmetry},
so strong symmetry implies weak symmetry.
For pure states, the converse also holds,
and the distinction arises only for mixed states.

For example, consider a qubit with the $\bb{Z}_2$ symmetry $U(g)=\sigma^z$ and
\begin{equation}
  \rho
  =
  \frac12
  \left(
  \ket{0}\!\bra{0}
  +
  \ket{1}\!\bra{1}
  \right).
\end{equation}
This state is weak symmetric because $\sigma^z\rho\sigma^z=\rho$,
but it is not strong symmetric because it mixes the two $\bb{Z}_2$ charge sectors.
Accordingly,
\begin{equation}
  \widetilde{L}_\rho(g)=0,
  \qquad
  L_\rho(g)=+\infty.
\end{equation}
The fidelity-based LCF therefore correctly identifies the state as symmetric in the sense used throughout the main text,
whereas the trace-based LCF remains sensitive to the mixture of different charges.

More generally, for an Abelian symmetry,
a weak symmetric state with charge probabilities $p_q$ satisfies
\begin{equation}
  \tr\!\bigl(\rho U(g)\bigr)
  =
  \sum_q p_q\,\mathrm{e}^{i\theta_q(g)}.
\end{equation}
The trace overlap can thus become small or vanish through cancellations between different charge sectors,
even when the density matrix is invariant under conjugation.
This explains both the divergence in the example above and the spike-like behavior of the trace-based LCF in Section~\ref{subsec:xxz_trace_compare}.

All results in the main text concerning symmetry breaking, symmetry restoration, entanglement asymmetry, and the quantum Mpemba effect use weak symmetry.
Strong symmetry is introduced only to clarify the interpretation of the trace-based LCF:
it is not a faithful measure of the ordinary mixed-state symmetry breaking studied here,
but it can be interpreted within a resource theory in which mixing different charge sectors is itself regarded as a resource \cite{Kusuki2026}.
This interpretation is also consistent with the upper bound
$\widetilde{L}_\rho(g)\leq L_\rho(g)$ derived in Eq.~\eqref{eq:LCFbound}.

\section{LCF of Fermionic Gaussian States}
\label{app:fermionic_gaussian_fidelity}

Here, we give the derivation of Eq.~\eqref{eq:LCF_trace}. Since $\rho_{A,g}$ in Sec.~\ref{subsec:cluster_gaussian} are Gaussian and conserve fermion number, they can be written in terms of the two-point correlation matrices as~\cite{peschel2003calculation}
\begin{equation}
  \rho_{A,g}
  =
  \det\!\left(\frac{I+\Gamma^{(g)}}{2}\right)
  \exp\!\left[
    \boldsymbol{\Psi}_A^\dagger
    \log R^{(g)}
    \boldsymbol{\Psi}_A
  \right],
  \label{eq:gaussian_density_matrix}
\end{equation}
where
\begin{equation}
  \boldsymbol{\Psi}_A
  :=
  \left({\bf \Psi}_1^\top,\ldots,{\bf \Psi}_\ell^\top\right)^\top,
  \quad
  R^{(g)}
  :=
  \left(I-\Gamma^{(g)}\right)
  \left(I+\Gamma^{(g)}\right)^{-1},
  \quad g\in\{1,{\rm e}\}.
  \label{eq:gaussian_R_definition}
\end{equation}
Using the equality
\begin{equation}
  \left[
    \boldsymbol{\Psi}_A^\dagger M\boldsymbol{\Psi}_A,
    \boldsymbol{\Psi}_A^\dagger N\boldsymbol{\Psi}_A
  \right]
  =
  \boldsymbol{\Psi}_A^\dagger[M,N]\boldsymbol{\Psi}_A,
  \label{eq:gaussian_bilinear_algebra}
\end{equation}
for matrices $M$ and $N$ and applying the Baker--Campbell--Hausdorff formula, the Uhlmann fidelity between the Gaussian states can be expressed as 
\begin{align}
  \tr
  \sqrt{\sqrt{\rho_{A,1}}\,\rho_{A,{\rm e}}\,\sqrt{\rho_{A,1}}}
  =
  \det\!\left(\frac{I+\Gamma^{(1)}}{2}\right)
  \tr
  \exp\left(
    \boldsymbol{\Psi}_A^\dagger
    \log\left(
      \sqrt{\sqrt{R^{(1)}}
      R^{({\rm e})}
      \sqrt{R^{(1)}}}
    \right)
    \boldsymbol{\Psi}_A
  \right).
  \label{eq:gaussian_sandwiched_state}
\end{align}

It remains to take the trace in the definition of the fidelity. For a Hermitian $2\ell\times2\ell$ matrix $K$ with eigenvalues $\xi_\alpha$, diagonalizing $K$ and summing over the empty and occupied states of each fermionic mode gives
\begin{equation}
  \mathrm{Tr}
  \exp\!\left(
    \boldsymbol{\Psi}_A^\dagger K\boldsymbol{\Psi}_A
  \right)
  =
  \prod_{\alpha=1}^{2\ell}\left(1+e^{\xi_\alpha}\right)
  =
  \det\!\left(I+e^K\right). 
  \label{eq:gaussian_trace_identity}
\end{equation}
Applying Eq.~\eqref{eq:gaussian_trace_identity} to Eq.~\eqref{eq:gaussian_sandwiched_state}, we obtain
\begin{align}
  F\!\left(\rho_{A,1},\rho_{A,{\rm e}}\right)
  =
  \det\!\left[
    \left(\frac{I+\Gamma^{(1)}}{2}\right)
    \left(
      I+
      \sqrt{
        \sqrt{R^{(1)}}
        R^{({\rm e})}
        \sqrt{R^{(1)}}
      }
    \right)
  \right].
  \label{eq:gaussian_fidelity_determinant}
\end{align}
Taking the negative logarithm of Eq.~\eqref{eq:gaussian_fidelity_determinant} and using $\log\det M=\mathrm{Tr}\log M$, we arrive at Eq.~\eqref{eq:LCF_trace}. 

\section{LCF in the Hydrodynamic Limit}
\label{app:cluster_lcf_hydrodynamics}
Here, we give the detailed derivation of Eq.~\eqref{eq:LCF_analytic}. We first expand the right-hand side of Eq.~\eqref{eq:LCF_trace} in terms of the moments of $\Gamma^{(1)}$ as
\begin{align}
  \widetilde{L}_A^{({\rm e})}(t)
  ={}&
  \sum_{m=1}^{\infty}
  \sum_{\substack{x_1,\ldots,x_m\in \mathbb{Z}_{\geq 1}\\y_1,\ldots,y_m\in\{0,1\}}}
  c_{\boldsymbol{x},\boldsymbol{y}}\,
  \mathrm{Tr}\!\left[
    \prod_{a=1}^{m}
    \left(\Gamma^{(1)}(t)\right)^{x_a}S^{y_a}
  \right],
  \label{eq:cluster_lcf_trace_expansion}
\end{align}
where $S:=I_\ell\otimes\sigma^z$ with $I_\ell$ being the $\ell\times\ell$ identity matrix and $c_{\boldsymbol{x},\boldsymbol{y}}$ are the expansion coefficients such that 
\begin{align}
  \sum_{m=1}^\infty 
  \sum_{\boldsymbol{x},\boldsymbol{y}}
  c_{\boldsymbol{x},\boldsymbol{y}}
    \mathrm{Tr}\!\left[
    \prod_{a=1}^{m}
    A^{x_a}B^{y_a}
    \right]
  = 
  -\mathrm{Tr}\log\left[ 
    \left(\frac{I+A}{2}\right)
    \left(
      I+\sqrt{\sqrt{\frac{I-A}{I+A}}\frac{I-BAB}{I+BAB}\sqrt{\frac{I-A}{I+A}}}
    \right)
  \right], \label{eq:sum_c}
 \end{align}
 for matrices $A$ and $B$. 
Substituting Eq.~\eqref{eq:cluster_toeplitz_matrix} into each trace on the right-hand side of Eq.~\eqref{eq:cluster_lcf_trace_expansion}, we obtain
\begin{align}
  &\mathrm{Tr}\!\left[
    \prod_{a=1}^{m}
    \left(\Gamma^{(1)}(t)\right)^{x_a}S^{y_a}
  \right]
  \nonumber\\
  &\quad=
  \int_{[-\pi,\pi]^M}
  \prod_{j=1}^{M}\frac{dk_j}{2\pi}
  \sum_{i_1=1}^{\ell}\cdots\sum_{i_M=1}^{\ell}
  \exp\!\left[
    i\sum_{j=1}^{M}k_j(i_j-i_{j+1})
  \right]
  \mathcal{W}(k_1,\ldots,k_M;t),
  \label{eq:cluster_trace_momentum_expansion}
\end{align}
where $M:=\sum_{a=1}^{m}x_a$, $i_{M+1}:=i_1$, and
\begin{align}
  \mathcal{W}(k_1,\ldots,k_M;t)
  =
  \tr\left[
    \prod_{a=1}^{m}
    \left(
      \prod_{j=X_{a-1}+1}^{X_a}
      \mathcal{T}^{(1)}(k_j,t)
    \right)
    (\sigma^z)^{y_a}
  \right],
  \label{eq:cluster_internal_trace}
\end{align}
with $X_a:=\sum_{b=1}^{a}x_b$. The summations over $i_j$ in Eq.~\eqref{eq:cluster_trace_momentum_expansion} can be replaced by integrals via the equality 
\begin{equation}
  \sum_{i=1}^{\ell}e^{iqi}
  =
  e^{iq(\ell+1)/2}
  \frac{\ell}{2}
  \int_{-1}^{1}du\,
  \frac{q}{2\sin(q/2)}
  e^{i\ell qu/2},
  \label{eq:cluster_finite_sum_integral}
\end{equation}
as
\begin{align}
  \mathrm{Tr}\left[
    \prod_{a=1}^{m}
    \left(\Gamma^{(1)}(t)\right)^{x_a}S^{y_a}
  \right]
  =
  \left(\frac{\ell}{4\pi}\right)^M
  \hspace{-10pt} 
  \int \limits_{[-\pi,\pi]^M} \hspace{-10pt}d^M \boldsymbol{k}
  \hspace{-10pt} 
  \int \limits_{[-1,1]^M} \hspace{-10pt} d^M \boldsymbol{u}
  \mathcal{C}(\boldsymbol{k}) 
  \mathcal{W}(\boldsymbol{k},t)
  e^{\frac{i\ell}{2}\sum_{j=1}^{M}k_j(u_j-u_{j+1})}
  ,
  \label{eq:cluster_trace_u_integral}
\end{align}
where $u_{M+1}:=u_1$, $k_0:=k_M$, and
\begin{equation}
  \mathcal{C}(k_1,\ldots,k_M)
  :=
  \prod_{j=1}^{M}
  \frac{k_j-k_{j-1}}
  {2\sin[(k_j-k_{j-1})/2]}. 
  \label{eq:cluster_C_function}
\end{equation}
Introducing the new variables
\begin{align}
  z_j = \begin{cases}
  u_1 & j=0 \\ 
  u_{j+1}-u_j & j=1,2,3,...,M-1
  \end{cases}, 
\end{align}
and integrating over $z_0$, Eq.~\eqref{eq:cluster_trace_u_integral} reduces to 
\begin{align}
  \mathrm{Tr}\!\left[
    \prod_{a=1}^{m}
    \left(\Gamma^{(1)}(t)\right)^{x_a}S^{y_a}
  \right]
  =
  \left(\frac{\ell}{4\pi}\right)^M
  \hspace{-10pt}\int \limits_{[-\pi,\pi]^M}\hspace{-10pt}d^M \boldsymbol{k}
  \hspace{-5pt}
  \int \limits_{\mathbb{R}^{M-1}}
  \hspace{-10pt}d^{M-1} \boldsymbol{z}
  \mu(\boldsymbol{z})
  \mathcal{C}(\boldsymbol{k})
  e^{-\frac{i\ell}{2}\sum_{j=1}^{M-1}z_j(k_j-k_M)}
  \mathcal{W}(\boldsymbol{k},t),
  \label{eq:cluster_trace_z_integral}
\end{align}
where 
\begin{equation}
  \mu(\boldsymbol{z})
  :=
  \max\!\left[
    0,
    2-
    \max_{0\leq q<M}\sum_{j=1}^{q}z_j
    +\min_{0\leq q<M}\sum_{j=1}^{q}z_j
  \right].
  \label{eq:cluster_mu_function}
\end{equation}
The function $\mu(\boldsymbol z)$ restricts the $\boldsymbol z$ integration to the region corresponding to $u_j\in[-1,1]$.

The Toeplitz symbol in $\mathcal{W}(\boldsymbol{k},t)$ can be expanded as 
\begin{align}
  \mathcal{T}^{(1)}(k,t)
  =
  \cos\theta_k\,\sigma^z
  +\sin\theta_k e^{-i\phi_k}
  e^{-2it\varepsilon_k}\sigma^+
  +\sin\theta_k e^{i\phi_k}
  e^{2it\varepsilon_k}\sigma^-,
  \label{eq:cluster_symbol_phase_decomposition}
\end{align}
where $\sigma^\pm:=(\sigma^x\pm i\sigma^y)/2$. Substituting Eq.~\eqref{eq:cluster_symbol_phase_decomposition} into Eq.~\eqref{eq:cluster_internal_trace} and grouping the terms according to the positions of $\sigma^\pm$, we obtain
\begin{align}
  \mathcal W(\boldsymbol k,t)
  ={}&
  \sum_{p=0}^{\lfloor M/2\rfloor}
  \sum_{1\leq j_1<\cdots<j_{2p}\leq M}
    \mathcal W_+(\boldsymbol k;j_1,\ldots,j_{2p})
    \exp\!\left(
      2it\sum_{a=1}^{2p}(-1)^a\varepsilon_{k_{j_a}}
    \right)
  \nonumber\\
  &+
  \sum_{p=1}^{\lfloor M/2\rfloor}
  \sum_{1\leq j_1<\cdots<j_{2p}\leq M}
    \mathcal W_-(\boldsymbol k;j_1,\ldots,j_{2p})
    \exp\!\left(
      -2it\sum_{a=1}^{2p}(-1)^a\varepsilon_{k_{j_a}}
    \right),
  \label{eq:cluster_alternating_phase}
\end{align}
where $\mathcal W_\pm$ are obtained from Eq.~\eqref{eq:cluster_internal_trace} through the replacements
\begin{align}
  \mathcal T^{(1)}(k_j,t)
  \longrightarrow
  \begin{cases}
    \sin\theta_{k_j}e^{\mp i\phi_{k_j}}\sigma^\pm,
    &j=j_a\text{ with odd }a,\\
    \sin\theta_{k_j}e^{\pm i\phi_{k_j}}\sigma^\mp,
    &j=j_a\text{ with even }a,\\
    \cos\theta_{k_j}\,\sigma^z,
    &j\notin\{j_1,\ldots,j_{2p}\},
  \end{cases}
\end{align}
with the upper and lower signs corresponding to $\mathcal W_+$ and $\mathcal W_-$, respectively.

Substituting Eq.~\eqref{eq:cluster_alternating_phase} into Eq.~\eqref{eq:cluster_trace_z_integral} and setting $t=\zeta\ell$, we obtain
\begin{multline}
  \mathrm{Tr} \left[
    \prod_{a=1}^{m}
    \left(\Gamma^{(1)}(t)\right)^{x_a}S^{y_a}
  \right]
  =
  \left(\frac{\ell}{4\pi}\right)^M
  \int_{[-\pi,\pi]^M}d^M\boldsymbol{k}
  \int_{\mathbb R^{M-1}}d^{M-1}\boldsymbol{z}\,
  \mu(\boldsymbol z)\mathcal C(\boldsymbol k)
  \\
  \times
  \Bigg\{
    \sum_{p=0}^{\lfloor M/2\rfloor}
    \sum_{1\leq j_1<\cdots<j_{2p}\leq M}
      \mathcal W_+(\boldsymbol k;j_1,\ldots,j_{2p})e^{i\ell\Phi_+}
  \\
  +
    \sum_{p=1}^{\lfloor M/2\rfloor}
    \sum_{1\leq j_1<\cdots<j_{2p}\leq M}
      \mathcal W_-(\boldsymbol k;j_1,\ldots,j_{2p})e^{i\ell\Phi_-}
  \Bigg\},
  \label{eq:cluster_trace_phase_expansion}
\end{multline}
where 
\begin{equation}
  \Phi_\pm(\boldsymbol{k};j_1,...,j_{2p})
  =
  -\frac12\sum_{j=1}^{M-1}z_j(k_j-k_M)
  \pm2\zeta\sum_{a=1}^{2p}(-1)^a
  \varepsilon_{k_{j_a}}.
  \label{eq:cluster_stationary_phase}
\end{equation}

We can evaluate the integrals in Eq.~\eqref{eq:cluster_trace_phase_expansion} for $\ell\gg1$ by applying the multidimensional stationary-phase formula: 
\begin{equation}
  \int_\Omega d^N\bm{\xi}\,
  f(\bm{\xi})e^{i\ell\Phi(\bm{\xi})}
  =
  \left(\frac{2\pi}{\ell}\right)^{N/2}
  \sum_{\bar{\bm{\xi}}:\,\nabla\Phi(\bar{\bm{\xi}})=0}
  \frac{
    f(\bar{\bm{\xi}})
    e^{i\ell\Phi(\bar{\bm{\xi}})
    +i\pi\,\operatorname{sgn}H_\Phi(\bar{\bm{\xi}})/4}
  }{
    \sqrt{\left|\det H_\Phi(\bar{\bm{\xi}})\right|}
  }
  +O\!\left(\ell^{-N/2-1}\right),
  \label{eq:multidimensional_stationary_phase}
\end{equation}
where $\Omega\subset\mathbb R^N$, $f$ is smooth and compactly supported in $\Omega$, $\Phi$ is real with isolated nondegenerate stationary points, $H_\Phi$ is its Hessian, and $\operatorname{sgn}H_\Phi$ is its signature. We identify each parameter in Eq.~\eqref{eq:multidimensional_stationary_phase} as 
\begin{align}
    k&=k_M,\\
    \Omega&=[-\pi,\pi]^{M-1}\times\mathbb R^{M-1},\\
    N&=2(M-1),\\
    \bm{\xi}&=(k_1,\ldots,k_{M-1},z_1,\ldots,z_{M-1}),\\
    \Phi&=\Phi_\pm,\\
    f(\boldsymbol{\xi})
    &=\left(\frac{\ell}{4\pi}\right)^M
    \mu(\boldsymbol{z})\mathcal{C}(\boldsymbol{k})
    \mathcal{W}_\pm(\boldsymbol{k};j_1,\ldots,j_{2p}).
    \label{eq:f}
\end{align}
The stationary conditions with respect to $z_j$ give
\begin{equation}
  \bar{k}_1=\cdots=\bar{k}_{M-1}=k,
  \label{eq:cluster_stationary_momenta}
\end{equation}
where the bars denote the values at a stationary point. The conditions with respect to $k_j$ give
\begin{equation}
  \bar{z}_j
  =
  \begin{cases}
    \displaystyle
    \pm4\zeta(-1)^a\varepsilon_k',
    &j=j_a\text{ for some }a,\\[3pt]
    0,
    &j\notin\{j_1,\ldots,j_{2p}\},
  \end{cases},
  \label{eq:cluster_stationary_positions}
\end{equation}
where $\varepsilon_k':=\partial_k\varepsilon_k$.
At the stationary points, Eq.~\eqref{eq:f} reduces to 
\begin{align}
  f(\bar{\boldsymbol{\xi}})
  =
  2\left(\frac{\ell}{4\pi}\right)^M
  \begin{cases}
    \mathcal{W}_0(k),
    &p=0,\\[3pt]
    \mathcal{W}_\pm(\bar{\boldsymbol{k}};j_1,\ldots,j_{2p})\max\!\left(1-2|\varepsilon_k'|\zeta,0\right),
    &p\geq1.
  \end{cases}
  \label{eq:cluster_stationary_prefactor}
\end{align}
Here, $\mathcal{W}_0(k)$ is obtained by replacing $\mathcal{T}^{(1)}(k,t)$ with $\cos \theta_k \sigma^z$. 
On the other hand, the Hessian of $\Phi$ at the stationary points reads 
\begin{equation}
  H_\Phi(\bar{\boldsymbol{\xi}})
  =
  \frac12
  \begin{pmatrix}
    J_\Phi & -I\\
    -I & 0
  \end{pmatrix},
  \label{eq:cluster_stationary_hessian}
\end{equation}
where $(J_\Phi)_{ij}:=2\partial_{k_i}\partial_{k_j}\Phi$ at the stationary point. If we denote as $\nu$ the eigenvalues of $J_\Phi$, the eigenvalues of $H_\Phi$ are given by $(\nu\pm\sqrt{\nu^2+4})/4$. We thus obtain 
\begin{equation}
  \mathrm{sgn}H_\Phi(\bar{\boldsymbol{\xi}})=0,\quad |\det H_\Phi(\bar{\boldsymbol{\xi}})|=4^{1-M}.
  \label{eq:cluster_hessian_determinant}
\end{equation}
Substituting Eqs.~\eqref{eq:cluster_stationary_prefactor} and \eqref{eq:cluster_hessian_determinant} into Eq.~\eqref{eq:multidimensional_stationary_phase} and summing over $p$, we obtain
\begin{multline}
  \mathrm{Tr}\left[
    \prod_{a=1}^{m}
    \left(\Gamma^{(1)}(t)\right)^{x_a}S^{y_a}
  \right]
  =
  \ell\int_{-\pi}^{\pi}\frac{dk}{2\pi}
  \max\!\left(1-2|\varepsilon_k'|\zeta,0\right)
  \operatorname{Tr} \left[
    \prod_{a=1}^{m}
    \mathcal{T}^{(1)}(k,0)^{x_a}(\sigma^z)^{y_a}
  \right]
  \\
  +
  \ell\int_{-\pi}^{\pi}\frac{dk}{2\pi}
  \min\!\left(2|\varepsilon_k'|\zeta,1\right)
  \operatorname{Tr}\!\left[
    \prod_{a=1}^{m}
    \left(\cos\theta_k\,\sigma^z\right)^{x_a}
    (\sigma^z)^{y_a}
  \right]
  +O(1).
  \label{eq:cluster_trace_word_asymptotic}
\end{multline}
Plugging it into \eqref{eq:cluster_lcf_trace_expansion} and taking the summation over $(\boldsymbol{x},\boldsymbol{y})$ using \eqref{eq:sum_c}, we arrive at Eq.~\eqref{eq:LCF_analytic}.

\clearpage
\bibliographystyle{JHEP}
\bibliography{main}

@article{travaglino2026gaussianasymmetrymeasure,
      title={A Gaussian asymmetry measure}, 
      author={Riccardo Travaglino and Pasquale Calabrese},
      year={2026},
      eprint={2604.26878},
      archivePrefix={arXiv},
      primaryClass={quant-ph},
      url={https://arxiv.org/abs/2604.26878}, 
}

@Article{Yamaguchi2023,
  author        = {Yamaguchi, Koji and Tajima, Hiroyasu},
  journal       = {Physical Review Letters},
  title         = {Beyond i.i.d. in the Resource Theory of Asymmetry: An Information-Spectrum Approach for Quantum Fisher Information},
  year          = {2023},
  month         = nov,
  number        = {20},
  pages         = {200203},
  volume        = {131},
  doi           = {10.1103/PhysRevLett.131.200203},
}

@article{Marvian_distillation,
  title={Coherence distillation machines are impossible in quantum thermodynamics},
  author={I. Marvian},
  journal={Nat. Commun.},
  volume={11},
  number={1},
  pages={25},
  year={2020},
  publisher={Nature Publishing Group},
  doi = {10.1038/s41467-019-13846-3}
}

@article{Marvian2022,
  title = {Operational Interpretation of Quantum Fisher Information in Quantum Thermodynamics},
  author = {Marvian, Iman},
  journal = {Phys. Rev. Lett.},
  volume = {129},
  issue = {19},
  pages = {190502},
  numpages = {6},
  year = {2022},
  month = {Oct},
  publisher = {American Physical Society},
  doi = {10.1103/PhysRevLett.129.190502},
  url = {https://link.aps.org/doi/10.1103/PhysRevLett.129.190502}
}

@phdthesis{Marvian_thesis,
  author={I. Marvian},
  title={Symmetry, Asymmetry and Quantum Information},
  school={the University of Waterloo},
  year={2012}
}

@article{Gour2008,
    title = {{The resource theory of quantum reference frames: Manipulations and monotones}},
    year = {2008},
    journal = {New Journal of Physics},
    author = {Gour, Gilad and Spekkens, Robert W.},
    number = {},
    pages = {033023},
    volume = {10},
    doi = {10.1088/1367-2630/10/3/033023},
    issn = {13672630},
    arxivId = {0711.0043}
}

@article{Marvian2013,
    title = {{The theory of manipulations of pure state asymmetry: I. Basic tools, equivalence classes and single copy transformations}},
    year = {2013},
    journal = {New Journal of Physics},
    author = {Marvian, Iman and Spekkens, Robert W.},
    pages = {033001},
    volume = {15},
    doi = {10.1088/1367-2630/15/3/033001},
    issn = {13672630}
}

@article{Marvian2014AsymmetryStates,
    title = {{Asymmetry properties of pure quantum states}},
    year = {2014},
    journal = {Physical Review A - Atomic, Molecular, and Optical Physics},
    author = {Marvian, Iman and Spekkens, Robert W.},
    number = {1},
    month = {7},
    volume = {90},
    publisher = {American Physical Society},
    doi = {10.1103/PhysRevA.90.014102},
    issn = {10941622},
    arxivId = {1105.1816}
}

@article{Yamaguchi2026,
  title = {Quantum Geometric Tensor Determines the Pure-State I.I.D. Conversion Rate in the Resource Theory of Asymmetry for Any Compact Lie Group},
  author = {Yamaguchi, Koji and Mitsuhashi, Yosuke and Shitara, Tomohiro and Tajima, Hiroyasu},
  journal = {Phys. Rev. X},
  volume = {16},
  issue = {3},
  pages = {031028},
  numpages = {75},
  year = {2026},
  month = {Aug},
  publisher = {American Physical Society},
  doi = {10.1103/qqf6-x85b},
  url = {https://link.aps.org/doi/10.1103/qqf6-x85b}
}

@book{Sagawa2022,
  title={Entropy, divergence, and majorization in classical and quantum thermodynamics},
  author={Sagawa, Takahiro},
  year={2022},
  publisher={Springer Nature}
}

@article{Gour2015,
title = {The resource theory of informational nonequilibrium in thermodynamics},
journal = {Physics Reports},
volume = {583},
pages = {1-58},
year = {2015},
note = {The resource theory of informational nonequilibrium in thermodynamics},
issn = {0370-1573},
doi = {10.1016/j.physrep.2015.04.003},
url = {https://www.sciencedirect.com/science/article/pii/S037015731500229X},
author = {Gilad Gour and Markus P. Müller and Varun Narasimhachar and Robert W. Spekkens and Nicole {Yunger Halpern}}
}

@article{vidal2003,
  title = {Efficient Classical Simulation of Slightly Entangled Quantum Computations},
  author = {Vidal, Guifr\'e},
  journal = {Phys. Rev. Lett.},
  volume = {91},
  issue = {14},
  pages = {147902},
  numpages = {4},
  year = {2003},
  month = {Oct},
  publisher = {American Physical Society},
  doi = {10.1103/PhysRevLett.91.147902},
  url = {https://link.aps.org/doi/10.1103/PhysRevLett.91.147902},
  eprint = "quant-ph/0301063",
  archivePrefix = "arXiv",
  primaryClass = "quant-ph"
}

@article{peschel2003calculation,
  title={Calculation of reduced density matrices from correlation functions},
  author={Peschel, Ingo},
  journal={Journal of Physics A: Mathematical and General},
  volume={36},
  number={14},
  pages={L205},
  year={2003},
  publisher={IOP Publishing},
url={https://iopscience.iop.org/article/10.1088/0305-4470/36/14/101/meta},
eprint = "cond-mat/0212631",
  archivePrefix = "arXiv",
  primaryClass = "cond-mat"
}

@article{hara2026,
  title = {Dynamics of entanglement asymmetry for space-inversion symmetry of free fermions on honeycomb lattices},
  author = {Hara, Ryogo and Endo, Shimpei and Yamashika, Shion},
  journal = {Phys. Rev. B},
  volume = {113},
  issue = {14},
  pages = {144313},
  numpages = {13},
  year = {2026},
  month = {Apr},
  publisher = {American Physical Society},
  doi = {10.1103/lpz6-3v48},
  url = {https://link.aps.org/doi/10.1103/lpz6-3v48},
  eprint = "2511.14114",
  archivePrefix = "arXiv",
  primaryClass = "cond-mat.quant-gas"
}

@article{yamashika2026_long_range,
  title = {Quantum Mpemba Effect in Long-Range Spin Systems},
  author = {Yamashika, Shion and Ares, Filiberto},
  journal = {Phys. Rev. Lett.},
  volume = {136},
  issue = {9},
  pages = {090402},
  numpages = {7},
  year = {2026},
  month = {Mar},
  publisher = {American Physical Society},
  doi = {10.1103/52y5-8kl2},
  url = {https://link.aps.org/doi/10.1103/52y5-8kl2},
  eprint = "2507.06636",
  archivePrefix = "arXiv",
  primaryClass = "cond-mat.stat-mech"
}

@article{Calabrese2005,
  title = {{Evolution of entanglement entropy in one-dimensional systems}},
  volume = {2005},
  ISSN = {1742-5468},
  url = {http://dx.doi.org/10.1088/1742-5468/2005/04/P04010},
  DOI = {10.1088/1742-5468/2005/04/p04010},
  number = {04},
  journal = {Journal of Statistical Mechanics: Theory and Experiment},
  publisher = {IOP Publishing},
  author = {Calabrese,  Pasquale and Cardy,  John},
  year = {2005},
  month = Apr,
  pages = {P04010},
  eprint = "cond-mat/0503393",
  archivePrefix = "arXiv",
  primaryClass = "cond-mat.stat-mech",
}

@Article{Ares2022,
  author        = {Ares, Filiberto and Murciano, Sara and Calabrese, Pasquale},
  journal       = {Nature Communications 14, 2036 (2023)},
  title         = {Entanglement asymmetry as a probe of symmetry breaking},
  year          = {2022},
  issn          = {2041-1723},
  month         = apr,
  number        = {1},
  volume        = {14},
  archiveprefix = {arXiv},
  copyright     = {Creative Commons Attribution 4.0 International},
  date          = {2022-07-29},
  doi           = {10.1038/s41467-023-37747-8},
  eprint        = {2207.14693},
  primaryclass  = {cond-mat.stat-mech},
  publisher     = {Springer Science and Business Media LLC},
}

@Article{Ares2023,
  author        = {Ares, Filiberto and Murciano, Sara and Vernier, Eric and Calabrese, Pasquale},
  journal       = {SciPost Phys. 15, 089 (2023)},
  title         = {Lack of symmetry restoration after a quantum quench: an entanglement asymmetry study},
  year          = {2023},
  issn          = {2542-4653},
  month         = sep,
  number        = {3},
  volume        = {15},
  archiveprefix = {arXiv},
  copyright     = {arXiv.org perpetual, non-exclusive license},
  date          = {2023-02-07},
  doi           = {10.21468/scipostphys.15.3.089},
  eprint        = {2302.03330},
  primaryclass  = {cond-mat.stat-mech},
  publisher     = {Stichting SciPost},
}

@Article{Ferro2023,
  author        = {Ferro, Florent and Ares, Filiberto and Calabrese, Pasquale},
  journal       = {J. Stat. Mech. (2024) 023101},
  title         = {Non-equilibrium entanglement asymmetry for discrete groups: the example of the XY spin chain},
  year          = {2023},
  issn          = {1742-5468},
  month         = feb,
  number        = {2},
  pages         = {023101},
  volume        = {2024},
  archiveprefix = {arXiv},
  copyright     = {arXiv.org perpetual, non-exclusive license},
  date          = {2023-07-13},
  doi           = {10.1088/1742-5468/ad138f},
  eprint        = {2307.06902},
  primaryclass  = {cond-mat.stat-mech},
  publisher     = {IOP Publishing},
}

@Article{Capizzi2023,
  author        = {Capizzi, Luca and Mazzoni, Michele},
  journal       = {JHEP 2023, 144 (2023)},
  title         = {Entanglement asymmetry in the ordered phase of many-body systems: the Ising Field Theory},
  year          = {2023},
  month         = jul,
  archiveprefix = {arXiv},
  copyright     = {arXiv.org perpetual, non-exclusive license},
  doi           = {10.48550/ARXIV.2307.12127},
  eprint        = {2307.12127},
  primaryclass  = {cond-mat.stat-mech},
  publisher     = {arXiv},
}

@Article{Capizzi2023a,
  author        = {Capizzi, Luca and Vitale, Vittorio},
  title         = {A universal formula for the entanglement asymmetry of matrix product states},
  year          = {2023},
  month         = oct,
  archiveprefix = {arXiv},
  copyright     = {arXiv.org perpetual, non-exclusive license},
  doi           = {10.48550/ARXIV.2310.01962},
  eprint        = {2310.01962},
  primaryclass  = {quant-ph},
  publisher     = {arXiv},
}

@Article{Fossati2024,
  author        = {Fossati, Michele and Ares, Filiberto and Dubail, Jerome and Calabrese, Pasquale},
  journal       = {JHEP 2024, 59 (2024)},
  title         = {Entanglement asymmetry in CFT and its relation to non-topological defects},
  year          = {2024},
  issn          = {1029-8479},
  month         = may,
  number        = {5},
  volume        = {2024},
  archiveprefix = {arXiv},
  copyright     = {arXiv.org perpetual, non-exclusive license},
  date          = {2024-02-05},
  doi           = {10.1007/jhep05(2024)059},
  eprint        = {2402.03446},
  primaryclass  = {hep-th},
  publisher     = {Springer Science and Business Media LLC},
}

@Article{Chen2023,
  author        = {Chen, Miao and Chen, Hui-Huang},
  title         = {Rényi entanglement asymmetry in 1+1-dimensional conformal field theories},
  year          = {2023},
  month         = oct,
  archiveprefix = {arXiv},
  copyright     = {Creative Commons Attribution 4.0 International},
  doi           = {10.48550/ARXIV.2310.15480},
  eprint        = {2310.15480},
  primaryclass  = {hep-th},
  publisher     = {arXiv},
}

@Article{Lastres2024,
  author        = {Lastres, Marco and Murciano, Sara and Ares, Filiberto and Calabrese, Pasquale},
  journal       = {J. Stat. Mech. (2025) 013107},
  title         = {Entanglement asymmetry in the critical XXZ spin chain},
  year          = {2024},
  issn          = {1742-5468},
  month         = jan,
  number        = {1},
  pages         = {013107},
  volume        = {2025},
  archiveprefix = {arXiv},
  copyright     = {arXiv.org perpetual, non-exclusive license},
  date          = {2024-07-08},
  doi           = {10.1088/1742-5468/ada497},
  eprint        = {2407.06427},
  primaryclass  = {cond-mat.stat-mech},
  publisher     = {IOP Publishing},
}

@Article{Fossati2024a,
  author        = {Fossati, Michele and Rylands, Colin and Calabrese, Pasquale},
  title         = {Entanglement asymmetry in CFT with boundary symmetry breaking},
  year          = {2024},
  month         = nov,
  archiveprefix = {arXiv},
  copyright     = {arXiv.org perpetual, non-exclusive license},
  doi           = {10.48550/ARXIV.2411.10244},
  eprint        = {2411.10244},
  primaryclass  = {hep-th},
  publisher     = {arXiv},
}

@Article{Kusuki2024,
  author        = {Kusuki, Yuya and Murciano, Sara and Ooguri, Hirosi and Pal, Sridip},
  title         = {Entanglement asymmetry and symmetry defects in boundary conformal field theory},
  year          = {2024},
  month         = nov,
  archiveprefix = {arXiv},
  copyright     = {arXiv.org perpetual, non-exclusive license},
  doi           = {10.48550/ARXIV.2411.09792},
  eprint        = {2411.09792},
  primaryclass  = {hep-th},
  publisher     = {arXiv},
}

@Article{Chen2024,
  author        = {Chen, Hui-Huang and Tang, Zi-Jun},
  title         = {Entanglement asymmetry in the Hayden-Preskill protocol},
  year          = {2024},
  month         = nov,
  archiveprefix = {arXiv},
  copyright     = {Creative Commons Attribution 4.0 International},
  doi           = {10.48550/ARXIV.2411.17695},
  eprint        = {2411.17695},
  primaryclass  = {hep-th},
  publisher     = {arXiv},
}

@Article{Ares2023a,
  author        = {Ares, Filiberto and Murciano, Sara and Piroli, Lorenzo and Calabrese, Pasquale},
  journal       = {Phys. Rev. D 110, L061901 (2024)},
  title         = {An entanglement asymmetry study of black hole radiation},
  year          = {2023},
  issn          = {2470-0029},
  month         = sep,
  number        = {6},
  pages         = {l061901},
  volume        = {110},
  archiveprefix = {arXiv},
  copyright     = {arXiv.org perpetual, non-exclusive license},
  date          = {2023-11-21},
  doi           = {10.1103/physrevd.110.l061901},
  eprint        = {2311.12683},
  primaryclass  = {hep-th},
  publisher     = {American Physical Society (APS)},
}

@Article{Russotto2024,
  author        = {Russotto, Angelo and Ares, Filiberto and Calabrese, Pasquale},
  journal       = {JHEP 06 (2025) 149},
  title         = {Non-Abelian entanglement asymmetry in random states},
  year          = {2024},
  issn          = {1029-8479},
  month         = jun,
  number        = {6},
  volume        = {2025},
  archiveprefix = {arXiv},
  copyright     = {arXiv.org perpetual, non-exclusive license},
  date          = {2024-11-20},
  doi           = {10.1007/jhep06(2025)149},
  eprint        = {2411.13337},
  primaryclass  = {hep-th},
  publisher     = {Springer Science and Business Media LLC},
}

@Article{Murciano2023,
  author        = {Murciano, Sara and Ares, Filiberto and Klich, Israel and Calabrese, Pasquale},
  journal       = {J. Stat. Mech. (2024) 013103},
  title         = {Entanglement asymmetry and quantum Mpemba effect in the XY spin chain},
  year          = {2023},
  issn          = {1742-5468},
  month         = jan,
  number        = {1},
  pages         = {013103},
  volume        = {2024},
  archiveprefix = {arXiv},
  copyright     = {arXiv.org perpetual, non-exclusive license},
  date          = {2023-10-11},
  doi           = {10.1088/1742-5468/ad17b4},
  eprint        = {2310.07513},
  primaryclass  = {cond-mat.stat-mech},
  publisher     = {IOP Publishing},
}

@Article{Ares2024,
  author        = {Ares, Filiberto and Vitale, Vittorio and Murciano, Sara},
  journal       = {Phys. Rev. B 111, 104312 (2025)},
  title         = {The quantum Mpemba effect in free-fermionic mixed states},
  year          = {2024},
  issn          = {2469-9969},
  month         = mar,
  number        = {10},
  pages         = {104312},
  volume        = {111},
  archiveprefix = {arXiv},
  copyright     = {arXiv.org perpetual, non-exclusive license},
  date          = {2024-05-14},
  doi           = {10.1103/physrevb.111.104312},
  eprint        = {2405.08913},
  primaryclass  = {cond-mat.stat-mech},
  publisher     = {American Physical Society (APS)},
}

@Article{Rylands2023,
  author        = {Rylands, Colin and Klobas, Katja and Ares, Filiberto and Calabrese, Pasquale and Murciano, Sara and Bertini, Bruno},
  journal       = {Phys. Rev. Lett. 133, 010401 (2024)},
  title         = {Microscopic origin of the quantum Mpemba effect in integrable systems},
  year          = {2023},
  issn          = {1079-7114},
  month         = jul,
  number        = {1},
  pages         = {010401},
  volume        = {133},
  archiveprefix = {arXiv},
  copyright     = {arXiv.org perpetual, non-exclusive license},
  date          = {2023-10-06},
  doi           = {10.1103/physrevlett.133.010401},
  eprint        = {2310.04419},
  primaryclass  = {cond-mat.stat-mech},
  publisher     = {American Physical Society (APS)},
}

@Article{Yamashika2024,
  author        = {Yamashika, Shion and Ares, Filiberto and Calabrese, Pasquale},
  journal       = {Phys. Rev. B 110, 085126 (2024)},
  title         = {Entanglement asymmetry and quantum Mpemba effect in two-dimensional free-fermion systems},
  year          = {2024},
  issn          = {2469-9969},
  month         = aug,
  number        = {8},
  pages         = {085126},
  volume        = {110},
  archiveprefix = {arXiv},
  copyright     = {arXiv.org perpetual, non-exclusive license},
  date          = {2024-03-07},
  doi           = {10.1103/physrevb.110.085126},
  eprint        = {2403.04486},
  primaryclass  = {cond-mat.stat-mech},
  publisher     = {American Physical Society (APS)},
}

@Article{Turkeshi2024,
  author        = {Turkeshi, Xhek and Calabrese, Pasquale and De Luca, Andrea},
  title         = {Quantum Mpemba Effect in Random Circuits},
  year          = {2024},
  month         = may,
  archiveprefix = {arXiv},
  copyright     = {Creative Commons Attribution 4.0 International},
  doi           = {10.48550/ARXIV.2405.14514},
  eprint        = {2405.14514},
  primaryclass  = {quant-ph},
  publisher     = {arXiv},
}

@Article{Liu2024,
  author        = {Liu, Shuo and Zhang, Hao-Kai and Yin, Shuai and Zhang, Shi-Xin},
  journal       = {Phys. Rev. Lett. 133, 140405 (2024)},
  title         = {Symmetry restoration and quantum Mpemba effect in symmetric random circuits},
  year          = {2024},
  issn          = {1079-7114},
  month         = oct,
  number        = {14},
  pages         = {140405},
  volume        = {133},
  archiveprefix = {arXiv},
  copyright     = {Creative Commons Attribution 4.0 International},
  date          = {2024-03-13},
  doi           = {10.1103/physrevlett.133.140405},
  eprint        = {2403.08459},
  primaryclass  = {quant-ph},
  publisher     = {American Physical Society (APS)},
}

@Article{Liu2024a,
  author        = {Liu, Shuo and Zhang, Hao-Kai and Yin, Shuai and Zhang, Shi-Xin and Yao, Hong},
  title         = {Quantum Mpemba effects in many-body localization systems},
  year          = {2024},
  month         = aug,
  archiveprefix = {arXiv},
  copyright     = {Creative Commons Attribution 4.0 International},
  doi           = {10.48550/ARXIV.2408.07750},
  eprint        = {2408.07750},
  primaryclass  = {cond-mat.dis-nn},
  publisher     = {arXiv},
}

@Article{Chalas2024,
  author        = {Chalas, Konstantinos and Ares, Filiberto and Rylands, Colin and Calabrese, Pasquale},
  journal       = {J. Stat. Mech. (2024) 103101},
  title         = {Multiple crossing during dynamical symmetry restoration and implications for the quantum Mpemba effect},
  year          = {2024},
  issn          = {1742-5468},
  month         = oct,
  number        = {10},
  pages         = {103101},
  volume        = {2024},
  archiveprefix = {arXiv},
  copyright     = {arXiv.org perpetual, non-exclusive license},
  date          = {2024-05-07},
  doi           = {10.1088/1742-5468/ad769c},
  eprint        = {2405.04436},
  primaryclass  = {cond-mat.stat-mech},
  publisher     = {IOP Publishing},
}

@Article{Caceffo2024,
  author        = {Caceffo, Fabio and Murciano, Sara and Alba, Vincenzo},
  journal       = {J. Stat. Mech. (2024) 063103},
  title         = {Entangled multiplets, asymmetry, and quantum Mpemba effect in dissipative systems},
  year          = {2024},
  issn          = {1742-5468},
  month         = jun,
  number        = {6},
  pages         = {063103},
  volume        = {2024},
  archiveprefix = {arXiv},
  copyright     = {arXiv.org perpetual, non-exclusive license},
  date          = {2024-02-05},
  doi           = {10.1088/1742-5468/ad4537},
  eprint        = {2402.02918},
  primaryclass  = {cond-mat.stat-mech},
  publisher     = {IOP Publishing},
}

@Article{Rylands2024,
  author        = {Rylands, Colin and Vernier, Eric and Calabrese, Pasquale},
  title         = {Dynamical symmetry restoration in the Heisenberg spin chain},
  year          = {2024},
  month         = sep,
  archiveprefix = {arXiv},
  copyright     = {Creative Commons Attribution 4.0 International},
  doi           = {10.48550/ARXIV.2409.08735},
  eprint        = {2409.08735},
  primaryclass  = {cond-mat.stat-mech},
  publisher     = {arXiv},
}

@Article{Yamashika2024a,
  author        = {Yamashika, Shion and Calabrese, Pasquale and Ares, Filiberto},
  journal       = {Physical Review A},
  title         = {Quenching from superfluid to free bosons in two dimensions: entanglement, symmetries, and quantum Mpemba effect},
  year          = {2024},
  issn          = {2469-9934},
  month         = apr,
  number        = {4},
  pages         = {043304},
  volume        = {111},
  archiveprefix = {arXiv},
  copyright     = {arXiv.org perpetual, non-exclusive license},
  date          = {2024-10-18},
  doi           = {10.1103/physreva.111.043304},
  eprint        = {2410.14299},
  primaryclass  = {cond-mat.stat-mech},
  publisher     = {American Physical Society (APS)},
}

@Article{Joshi2024,
  author        = {Joshi, Lata Kh and Franke, Johannes and Rath, Aniket and Ares, Filiberto and Murciano, Sara and Kranzl, Florian and Blatt, Rainer and Zoller, Peter and Vermersch, Benoît and Calabrese, Pasquale and Roos, Christian F. and Joshi, Manoj K.},
  journal       = {Phys. Rev. Lett. 133, 010402, 2024},
  title         = {Observing the quantum Mpemba effect in quantum simulations},
  year          = {2024},
  month         = jan,
  archiveprefix = {arXiv},
  copyright     = {arXiv.org perpetual, non-exclusive license},
  doi           = {10.48550/ARXIV.2401.04270},
  eprint        = {2401.04270},
  primaryclass  = {quant-ph},
  publisher     = {arXiv},
}

@Article{Gour2009,
  author        = {Gour, Gilad and Marvian, Iman and Spekkens, Robert W.},
  journal       = {Phys. Rev. A 80, 012307 (2009)},
  title         = {Measuring the quality of a quantum reference frame: the relative entropy of frameness},
  year          = {2009},
  issn          = {1094-1622},
  month         = jul,
  number        = {1},
  pages         = {012307},
  volume        = {80},
  archiveprefix = {arXiv},
  copyright     = {arXiv.org perpetual, non-exclusive license},
  date          = {2009-01-07},
  doi           = {10.1103/physreva.80.012307},
  eprint        = {0901.0943},
  primaryclass  = {quant-ph},
  publisher     = {American Physical Society (APS)},
}

@Article{Groot2021,
  author        = {de Groot, Caroline and Turzillo, Alex and Schuch, Norbert},
  journal       = {Quantum 6, 856 (2022)},
  title         = {Symmetry Protected Topological Order in Open Quantum Systems},
  year          = {2021},
  issn          = {2521-327X},
  month         = nov,
  pages         = {856},
  volume        = {6},
  archiveprefix = {arXiv},
  copyright     = {Creative Commons Attribution 4.0 International},
  date          = {2021-12-08},
  doi           = {10.22331/q-2022-11-10-856},
  eprint        = {2112.04483},
  primaryclass  = {quant-ph},
  publisher     = {Verein zur Forderung des Open Access Publizierens in den Quantenwissenschaften},
}

@Article{Lessa2024,
  author        = {Lessa, Leonardo A. and Ma, Ruochen and Zhang, Jian-Hao and Bi, Zhen and Cheng, Meng and Wang, Chong},
  journal       = {PRX Quantum 6, 010344 (2025)},
  title         = {Strong-to-Weak Spontaneous Symmetry Breaking in Mixed Quantum States},
  year          = {2024},
  issn          = {2691-3399},
  month         = mar,
  number        = {1},
  pages         = {010344},
  volume        = {6},
  archiveprefix = {arXiv},
  copyright     = {Creative Commons Attribution Non Commercial No Derivatives 4.0 International},
  date          = {2024-05-06},
  doi           = {10.1103/prxquantum.6.010344},
  eprint        = {2405.03639},
  primaryclass  = {quant-ph},
  publisher     = {American Physical Society (APS)},
}

@Article{Sala2024,
  author        = {Sala, Pablo and Gopalakrishnan, Sarang and Oshikawa, Masaki and You, Yizhi},
  title         = {Spontaneous Strong Symmetry Breaking in Open Systems: Purification Perspective},
  year          = {2024},
  month         = may,
  archiveprefix = {arXiv},
  copyright     = {arXiv.org perpetual, non-exclusive license},
  doi           = {10.48550/ARXIV.2405.02402},
  eprint        = {2405.02402},
  primaryclass  = {quant-ph},
  publisher     = {arXiv},
}

@Article{Kusuki2023,
  author        = {Kusuki, Yuya and Murciano, Sara and Ooguri, Hirosi and Pal, Sridip},
  journal       = {JHEP 2023, 216 (2023)},
  title         = {Symmetry-resolved Entanglement Entropy, Spectra \& Boundary Conformal Field Theory},
  year          = {2023},
  issn          = {1029-8479},
  month         = nov,
  number        = {11},
  volume        = {2023},
  archiveprefix = {arXiv},
  copyright     = {arXiv.org perpetual, non-exclusive license},
  date          = {2023-09-06},
  doi           = {10.1007/jhep11(2023)216},
  eprint        = {2309.03287},
  primaryclass  = {hep-th},
  publisher     = {Springer Science and Business Media LLC},
}

@Article{Calabrese2016,
  author        = {Calabrese, Pasquale and Cardy, John},
  journal       = {J. Stat. Mech. (2016) 064003},
  title         = {Quantum quenches in 1+1 dimensional conformal field theories},
  year          = {2016},
  issn          = {1742-5468},
  month         = jun,
  number        = {6},
  pages         = {064003},
  volume        = {2016},
  archiveprefix = {arXiv},
  copyright     = {arXiv.org perpetual, non-exclusive license},
  date          = {2016-03-09},
  doi           = {10.1088/1742-5468/2016/06/064003},
  eprint        = {1603.02889},
  primaryclass  = {cond-mat.stat-mech},
  publisher     = {IOP Publishing},
}

@Article{Benini2025,
  author        = {Benini, Francesco and Calabrese, Pasquale and Fossati, Michele and Singh, Amartya Harsh and Venuti, Marco},
  title         = {Entanglement Asymmetry for Higher and Noninvertible Symmetries},
  year          = {2025},
  month         = sep,
  archiveprefix = {arXiv},
  copyright     = {arXiv.org perpetual, non-exclusive license},
  doi           = {10.48550/ARXIV.2509.16311},
  eprint        = {2509.16311},
  primaryclass  = {hep-th},
  publisher     = {arXiv},
}

@Article{Yamashika2025,
  author        = {Yamashika, Shion and Endo, Shimpei and Tajima, Hiroyasu},
  title         = {Quantum Fisher Information as a Measure of Symmetry Breaking in Quantum Many-Body Systems},
  year          = {2025},
  month         = sep,
  archiveprefix = {arXiv},
  copyright     = {arXiv.org perpetual, non-exclusive license},
  doi           = {10.48550/ARXIV.2509.07468},
  eprint        = {2509.07468},
  primaryclass  = {cond-mat.stat-mech},
  publisher     = {arXiv},
}

@Article{Cardy2004,
  author        = {Cardy, John},
  journal       = {Encyclopedia of Mathematical Physics, (Elsevier, 2006)},
  title         = {Boundary Conformal Field Theory},
  year          = {2004},
  month         = nov,
  archiveprefix = {arXiv},
  copyright     = {Assumed arXiv.org perpetual, non-exclusive license to distribute this article for submissions made before January 2004},
  doi           = {10.48550/ARXIV.HEP-TH/0411189},
  eprint        = {hep-th/0411189},
  primaryclass  = {hep-th},
  publisher     = {arXiv},
}

@Article{Fujimura2025,
  author        = {Fujimura, Harunobu and Shimamori, Soichiro},
  journal       = {JHEP},
  title         = {Entanglement Asymmetry and Quantum {Mpemba} Effect for {Non-Abelian} Global Symmetry},
  year          = {2026},
  pages         = {244},
  volume        = {03},
  archiveprefix = {arXiv},
  doi           = {10.1007/JHEP03(2026)244},
  eprint        = {2509.05597},
  primaryclass  = {hep-th},
  reportnumber  = {OU-HET 1284},
}

@Article{Bonsignori2023,
  author        = {Bonsignori, Riccarda and Capizzi, Luca and Panopoulos, Pantelis},
  journal       = {JHEP},
  title         = {Boundary Symmetry Breaking in CFT and the string order parameter},
  year          = {2023},
  pages         = {027},
  volume        = {05},
  archiveprefix = {arXiv},
  doi           = {10.1007/JHEP05(2023)027},
  eprint        = {2301.08676},
  primaryclass  = {hep-th},
}

@InProceedings{Shitara2023,
  author        = {Shitara, Tomohiro and Mitsuhashi, Yosuke and Tajima, Hiroyasu},
  title         = {The i.i.d. State Convertibility in the Resource Theory of Asymmetry for Finite Groups},
  year          = {2023},
  month         = {12},
  archiveprefix = {arXiv},
  eprint        = {2312.15758},
  primaryclass  = {quant-ph},
}

@Article{Barad2024,
  author       = {Ruhanshi Barad and Qicheng Tang and Wei Zhu and Xueda Wen},
  title        = {Universal time evolution of string order parameter in quantum critical systems with boundary invertible or non-invertible symmetry breaking},
  date         = {2024-10-21},
  doi          = {10.1103/PhysRevB.111.165121},
  eprint       = {2410.16402},
  eprintclass  = {cond-mat.str-el},
  eprinttype   = {arXiv},
  journaltitle = {Phys. Rev. B 111, 165121 (2025)},
}

@Article{Kusuki2026,
  author        = {Kusuki, Yuya and Pal, Sridip and Tajima, Hiroyasu},
  title         = {Resource-Theoretic Quantifiers of Weak and Strong Symmetry Breaking: Strong Entanglement Asymmetry and Beyond},
  year          = {2026},
  month         = {1},
  archiveprefix = {arXiv},
  eprint        = {2601.20924},
  primaryclass  = {hep-th},
  reportnumber  = {KYUSHU-HET-344, RIKEN-iTHEMS-Report-26},
}

@Article{Vaccaro2008,
  author        = {Vaccaro, J. A. and Anselmi, F. and Wiseman, H. M. and Jacobs, K.},
  title         = {Tradeoff between extractable mechanical work, accessible entanglement, and ability to act as a reference system, under arbitrary superselection rules},
  journal       = {Phys. Rev. A},
  year          = {2008},
  volume        = {77},
  number        = {3},
  pages         = {032114},
  doi           = {10.1103/PhysRevA.77.032114},
  eprint        = {quant-ph/0501121},
  archiveprefix = {arXiv},
  primaryclass  = {quant-ph},
}

@Article{Benini:2024xjv,
  author        = {Benini, Francesco and Godet, Victor and Singh, Amartya Harsh},
  journal       = {Prog. Theor. Exp. Phys.},
  title         = {Entanglement asymmetry in conformal field theory and holography},
  year          = {2025},
  number        = {6},
  pages         = {063B05},
  volume        = {2025},
  archiveprefix = {arXiv},
  doi           = {10.1093/ptep/ptaf080},
  eprint        = {2407.07969},
  primaryclass  = {hep-th},
  reportnumber  = {SISSA 14/2024/FISI},
}

@Article{Florio2025,
  author        = {Florio, Adrien and Murciano, Sara},
  journal       = {Phys. Rev. D},
  title         = {Entanglement asymmetry in gauge theories: chiral anomaly in the finite temperature massless {Schwinger} model},
  year          = {2026},
  pages         = {L091901},
  volume        = {113},
  archiveprefix = {arXiv},
  doi           = {10.1103/q386-v4ch},
  eprint        = {2511.01966},
  primaryclass  = {hep-th},
}

@Article{Benini:2025hbj,
  author        = {Benini, Francesco and Garc{\'i}a-Valdecasas, Eduardo and Vitouladitis, Stathis},
  journal       = {JHEP},
  title         = {Higher-form entanglement asymmetry. {Part I.} The limits of symmetry breaking},
  year          = {2026},
  pages         = {202},
  volume        = {05},
  archiveprefix = {arXiv},
  doi           = {10.1007/JHEP05(2026)202},
  eprint        = {2512.15898},
  primaryclass  = {hep-th},
}

@Article{DiGiulio:2025ems,
  author        = {Di Giulio, Giuseppe and Turkeshi, Xhek and Murciano, Sara},
  journal       = {Entropy},
  title         = {Measurement-induced symmetry restoration and quantum {Mpemba} effect},
  year          = {2025},
  number        = {4},
  pages         = {407},
  volume        = {27},
  archiveprefix = {arXiv},
  doi           = {10.3390/e27040407},
  eprint        = {2502.19506},
  primaryclass  = {quant-ph},
}

@Article{Ares:2025onj,
  author        = {Ares, Filiberto and Calabrese, Pasquale and Murciano, Sara},
  journal       = {Nature Reviews Physics},
  title         = {The quantum {Mpemba} effects},
  year          = {2025},
  number        = {8},
  pages         = {451--460},
  volume        = {7},
  archiveprefix = {arXiv},
  doi           = {10.1038/s42254-025-00838-0},
  eprint        = {2502.08087},
  primaryclass  = {cond-mat.stat-mech},
}

@article{Fraenkel2020,
    author = "Fraenkel, Shachar and Goldstein, Moshe",
    title = "{Symmetry resolved entanglement: Exact results in 1D and beyond}",
    eprint = "1910.08459",
    archivePrefix = "arXiv",
    primaryClass = "cond-mat.stat-mech",
    doi = "10.1088/1742-5468/ab7753",
    journal = "J. Stat. Mech.",
    volume = "2003",
    number = "3",
    pages = "033106",
    year = "2020"
}

\end{document}